\documentclass[referee]{aa}  
\usepackage{graphicx}
\usepackage{dirtytalk}
\usepackage{txfonts}
\usepackage{multirow}
\usepackage{lineno}
\usepackage[colorlinks=false]{hyperref}
\begin{document}

   \title{Coronal mass ejections observed by heliospheric imagers at 0.2 and 1 au:}
   \subtitle{the events on April 1 and 2, 2019}
   \author{Carlos R. Braga
          \inst{1,2}
          \and
          Angelos Vourlidas\inst{2}
   }
   \institute{George Mason University, 
                4400 University Drive, Faixfax, VA, 22030, US\\
              \email{cbraga@gmu.edu}
         \and
             The Johns Hopkins University Applied Physics Laboratory, 
             Laurel, MD 20723, US
             }

   \date{Received September 22, 2020}


  \abstract
   {We study two coronal mass ejections (CMEs) observed between April 1-2, 2019 by both the inner Wide-Field Imager for Parker Solar Probe (WISPR-I) on board Parker Solar Probe spacecraft (located between about 46 and 38 solar radii during this period) and the inner heliospheric imager (HI-1) on board the Solar Terrestrial Relations Observatory Ahead spacecraft (STEREO-A), orbiting the Sun at about 0.96 au. This is the first study of CME observations from two viewpoints in similar directions but at considerably different solar distances.}
   {Our objective is deriving CME kinematics from WISPR-I observations and comparing them with results from HI-1. This allows us to understand how the PSP observations affect the CME kinematics, especially due to its proximity to the Sun.} 
   {We estimate the CME positions, speeds, accelerations, propagation directions and longitudinal deflections using imaging observations from two spacecraft, and a set of analytical expressions that consider the CME as a point structure and take into account the rapid change in spacecraft position. We derive the kinematics using each viewpoint independently and both viewpoints as a constraint.}
   {We find that both CMEs are slow ($< 400\ km\ s^{-1}$), propagating eastward of the Sun-Earth line (westward of PSP and STEREO-A). The second CME seems to accelerate between $\sim 0.1$ to $\sim 0.2\ au$ and deflect westward with an angular speed consistent with the solar rotation speed. We find some discrepancies in the CME solar distance (up to $0.05\ au$, particularly for CME \#1), latitude (up to $\sim10^{\circ}$) and longitude (up to $24^{\circ}$) when comparing results from different fit cases (different observations or set of free parameters).} 
   {Discrepancies in longitude are likely due to the feature tracked visually rather than instrumental biases or fit assumptions. For similar reasons, the CME \#1 solar distance, as derived from WISPR-I observations, is larger than the HI-1 result, regardless of the fit parameters considered. Error estimates for CME kinematics do not show any clear trend associated to the observing instrument. The source region location and the lack of any clear in situ counterparts (both at near-Earth and at PSP) support our estimate of the propagation direction for both events.}

    \titlerunning{CMEs on 2019 April observed by the PSP and STEREO-A} 
    \authorrunning{Braga \& Vourlidas}
    
   \maketitle

%

\section{Introduction}

\label{sec:intro}

Remote sensing of coronal mass ejections (CMEs) from space-borne instruments has been done from locations close to 1 au for over 4 decades \citep{Tousey1973, MacQueen1974,MacQueen1980,Sheeley1980}. Since 1996, observations from the Large-Angle and Spectrometric Coronagraph \citep[LASCO;][]{Brueckner1995} on board the Solar and Heliospheric Observatory \citep[SOHO;][]{Domingo1995} have been recording thousands of CMEs. In 2006, imaging observations from observatories far from the Sun-Earth line began with the launch of the twin Solar Terrestial Relations Observatory spacecraft \citep[STEREO;][]{Kaiser2007}. In the STEREO era, the coronagraphic and heliospheric observations have been undertaken by the Sun Earth Coronal Connection and Heliospheric Investigation instrument suite \citep[SECCHI;][]{Howard2008}, a suite of two white light coronagraphs, to heliospheric imagers and one extreme ultraviolet disk imager. All  observations above are performed from $\sim 1\ au$ locations. 

In 2018, the Parker Solar Probe mission \citep[PSP;][]{Fox2015a} was launched to become the first mankind object to enter a star's atmosphere. The spacecraft reached positions lower than 0.3 au and observed the solar corona through remote sensing from this region for the first time \citep{Howard2019}. It has a highly elliptical heliocentric orbit and is mainly Sun-pointed with aphelia between Venus and Earth. The spacecraft perihelion distance is lowered gradually, beginning with approximately 35 solar radii at the start of the mission (2018) reaching to less than 10 solar radii at the end of the 7-year mission prime phase \citep{Fox2015a, Howard2019}.

In addition to three in situ payloads, PSP includes an imaging telescope, the Wide-Field Imager for Solar Probe \citep[WISPR:[]{Vourlidas2015}. WISPR comprises two heliospheric imagers mounted on the spacecraft ram side, which allows observations of the solar corona ahead. 

Several methodologies exist for deriving the kinematics of CMEs from heliospheric imager observations. Most of them were developed for observations from STEREO, such as fixed-$\phi$ \citep{Sheeley1999,Kahler2007,Rouillard2008} and harmonic mean \citep{Lugaz2009}, to name a few. 
Recently, \citet{Liewer2019} used synthetic images to introduce a method to derive the CME kinematics in WISPR field of view (FOV). 

\citet{Howard2019} described the first CME observations from WISPR involving no kinematics. \cite{Wood2020} reconstructed the November 5, 2018 CME morphology considering a flux rope structure and using observations from LASCO/SOHO coronagraphs and WISPR. The authors also derived the CME kinematics, but only on LASCO coronagraphs. \cite{Hess2020} calculated kinematics of a CME observed on November 1, 2018 during the first PSP encounter using observations from both WISPR-I and LASCO. 

In this article we study two CMEs observed by WISPR during its second perihelion, on April 1 and 2, 2019. Hereafter, we refer them as CME \#1 and \#2. In this period, PSP was approaching its second perihelion, which occurred on April 4. Located approximately five times further away from the Sun along the same radial, STEREO-A/SECCHI also observed these events and provides us with a second viewpoint. 

The objective of this study is to investigate, for the first time, the CME kinematics using multi-viewpoint observations with heliospheric imagers at similar angular distance from the CME but at different solar distances. We aim to understand if the speed and direction of propagation derived using spacecraft at different distances are similar or if they change significantly because of some error source. 

The article is organized as follows: in Section \ref{sec:observations}, we briefly describe the available observations; the methodology to calculate the CME position, conceived especially for WISPR, is explained in Section \ref{sec:pos3d}; results and discussion are included in Section \ref{sec:results_and_discussion}; we also discuss in situ observations of these CMEs using measurements from PSP and from the Earth's vicinity in Section \ref{sec:geo}; finally, we summarize our findings in Section \ref{sec:final_remarks}.

\section{Observations}
\label{sec:observations}

\begin{figure*}[t]
\includegraphics[width=\hsize]{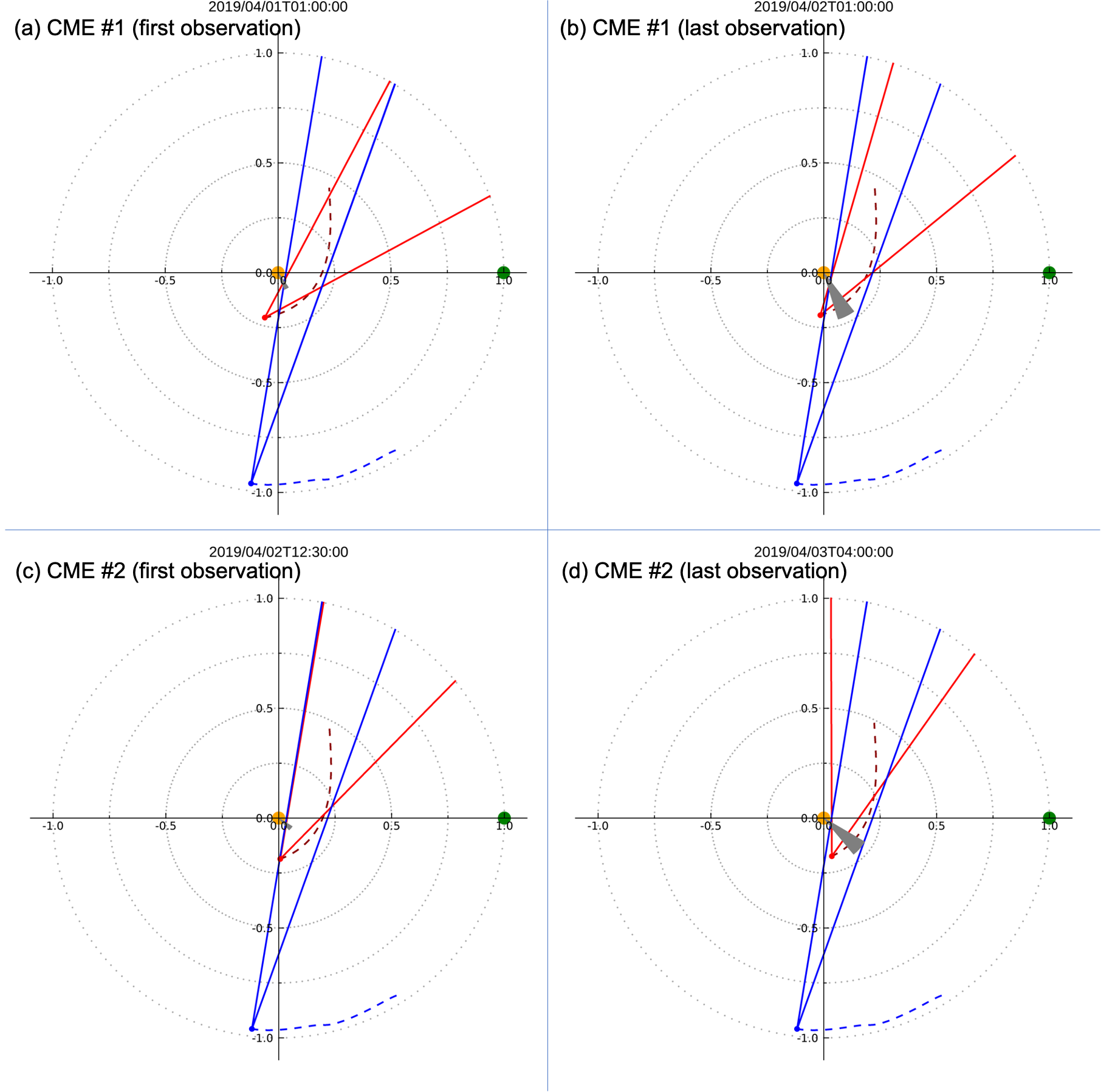}
\caption{Position of PSP (red circle) and STEREO-A (blue circle) in the ecliptic plane during the first CME (upper row) and second CME (lower row). 
The gray region represents CME \#1 (upper row) and CME \#2 (lower row) using positions calculated in Sections \ref{sec:results_cme1} and \ref{sec:results_cme2} (Tables \ref{tab:results_pos3d_cme1_acceleration_tied_position} and \ref{tab:results_pos3d_cme2_acceleration_tied_position}) and widths from Section \ref{sec:width}.
The Sun and Earth's positions are represented by the orange and green circles, respectively. The left panels (a and c) correspond to the first observation of each CME in WISPR-I FOV, and the right ones correspond to the last observation on HI-1. The approximate FOV of HI-1 and WISPR-I are represented by the red and blue lines, respectively. The dashed line represent the spacecraft position in an arbitrary future point in the orbit.} 
\label{fig:spacecraft_position_and_view}
\end{figure*}

In this study, we use observations from two heliospheric imagers, STEREO-A/HI-1 and PSP/WISPR-I. Both spacecraft are located at a similar angular distance from the Sun-Earth line: STEREO-A is at $\sim 97^{\circ}$ eastward from the Sun-Earth line, while the PSP location ranges from $106^{\circ}$ to $80^{\circ}$ between April 1 to 3, 2019, when the CMEs are within both FOVs. As for latitude, both spacecraft are slightly below the solar equatorial plane ($1\sim3^\circ$ for PSP and $2\sim3^\circ$ for STEREO-A) and close to the ecliptic plane (the PSP latitude ranges from $-0.3^{\circ}$ to $1.3^{\circ}$ and STEREO-A's is at $-0.1^{\circ}$). The fundamental difference in the STEREO-A - PSP configuration is their heliocentric distance: $\sim 0.96\ au$ for the former and $\sim 0.20\ au$ for the latter. We show the positions and FOVs of both imagers in Figure \ref{fig:spacecraft_position_and_view}.

The WISPR $95^{\circ}$ FOV is split across two telescopes: WISPR-I (inner; $13.5^{\circ}$ to $-53^{\circ}$ elongation) and WISPR-O (outer; $50.5^{\circ}$ to $-108.5^{\circ}$ elongation) \citep{Vourlidas2015}. For both CMEs studied here, observations from WISPR-O are excluded since the CMEs are outside their FOVs. HI-1 has a $20^{\circ}$ FOV centered at $14^{\circ}$ elongation along the ecliptic plane.

We inspected SECCHI/COR2 coronagraph observations in April 1-2, 2019. To our assessment, only two CMEs are observed and their timing is consistent with the CMEs studied here. As the two CMEs are not simultaneous, and we observe no other CMEs in the period, we can unambiguously link HI-1 and WISPR-I observations to each CME.

\begin{figure*}
\centering
\includegraphics[width=\hsize]{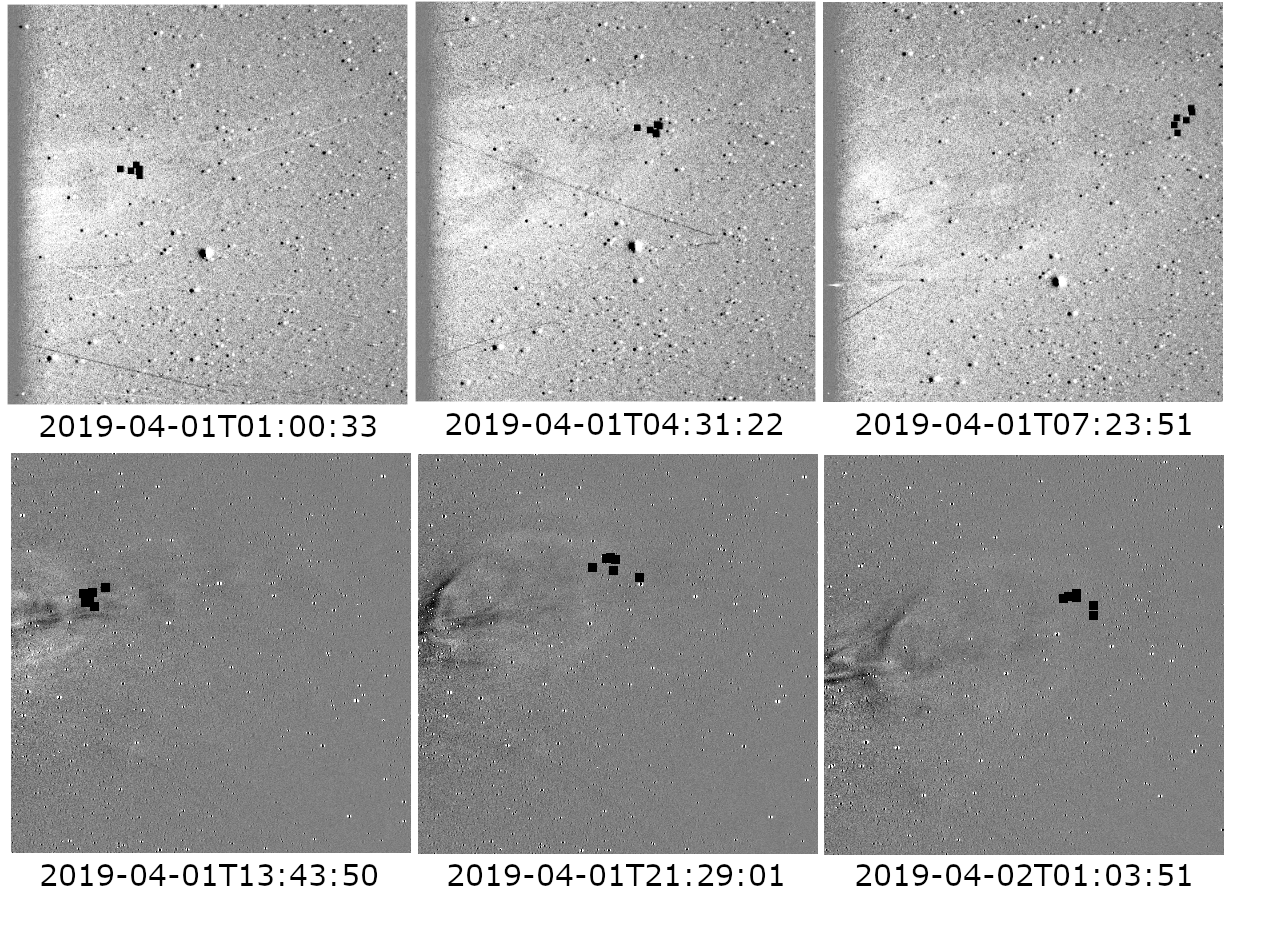}
\caption{The coronal mass ejections observed on April 1, 2019 by the WISPR-I (upper row) imager onboard the Parker Solar Probe and the HI-1 onboard STEREO-A (lower row). Here we display only 3 observations for each instrument, one close to first observation (left column), other close to the center of the observation period (center column) and other close to the last observation (right column). All images shown here are running differences cropped to evidence the CME region. The 6 black squares in each frame represent the CME front point selected in each visual inspection (see details in Section \ref{sec:pos3d}).
\label{fig:cme1}}
\end{figure*}

\begin{figure*}
\includegraphics[width=\hsize]{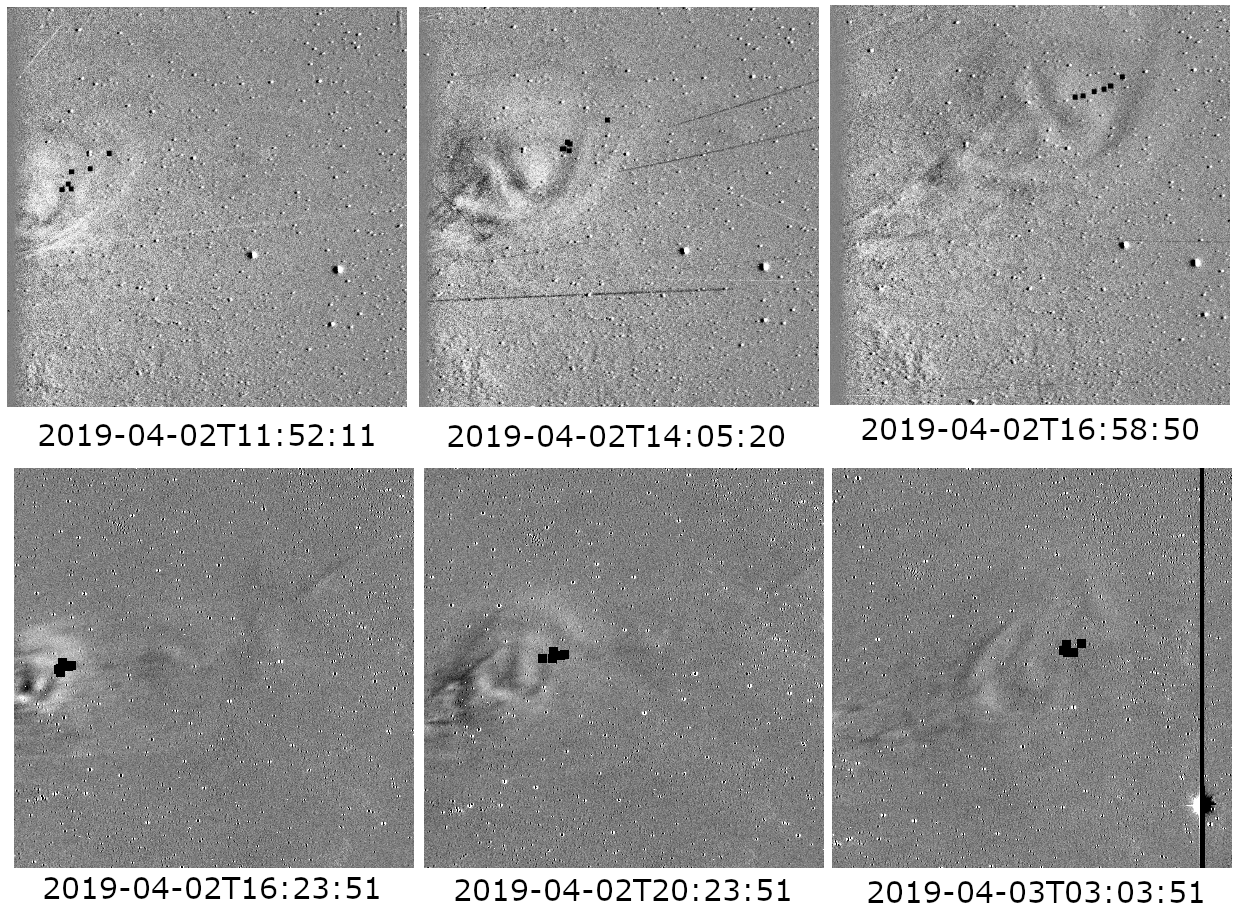}
\caption{The coronal mass ejections observed on April 2, 2019 by the WISPR-I (upper row) imager on board the Parker Solar Probe and the HI-1 on board STEREO-A (lower row). Here we display only 3 observations for each instrument, one close to first observation (left column), other close to the center of the observation period (center column) and other close to the last observation (right column). All images shown here are running differences cropped to evidence the CME region. The 6 black squares in each frame represent the CME front point selected in each visual inspection (see details in Section \ref{sec:pos3d}).
\label{fig:cme2}}
\end{figure*}

The CME \#1 and \#2 observations in the HI-1 and WISPR-I FOVs are shown in Figures \ref{fig:cme1} and \ref{fig:cme2}, respectively. In all images, the Sun is located outside the left edge of the FOV. 
The WISPR-I images are running difference images produced from calibrated data but without background removal (Level-2). We use HI-1 running difference images produced by \texttt{rdif\textunderscore hi.pro}\normalfont, an IDL procedure available in \texttt{SolarSoft} \normalfont \citep{Freeland1998}. The procedure reduces the star field in the running differences images. The normal running difference method results in bright and dark features that can compromise the visualization of the CME front. 

In WISPR-I, we followed the CME \#1 and \#2 fronts for approximately 6 and 5 hours, respectively. The WISPR-I cadence was approximately 20 minutes. In HI-1, we followed the fronts for 11 and 13 hours, respectively. The HI-1 cadence is 40 minutes.

\section{Calculating the CME position}
\label{sec:pos3d}

The PSP heliocentric longitude and distance change rapidly compared to previous missions such as STEREO. While a one-degree change in longitude takes days on STEREO, it occurs in less than a day for PSP, particularly close to perihelion. Many methodologies applied on heliospheric imagers from STEREO rely on assumptions about the spacecraft position variation during the typical CME observations course.  
For example, the CME position angle (the counterclockwise angle relative to solar north) remains constant. Therefore, it is customary to select a fixed position angle and track the feature along this line, building the so-called J-maps. However, as \citet{Liewer2019} showed using synthetic coronagraph images, radially moving structures, with a constant velocity, can change position angle depending on where they are located relative to the orbit plane. Therefore, a J-map of such a structure may not increase monotonically over time, as has been the case of J-maps built with STEREO observations. Thus, they cannot be directly applied to WISPR observations. 

\begin{figure}
\includegraphics[width=9cm]{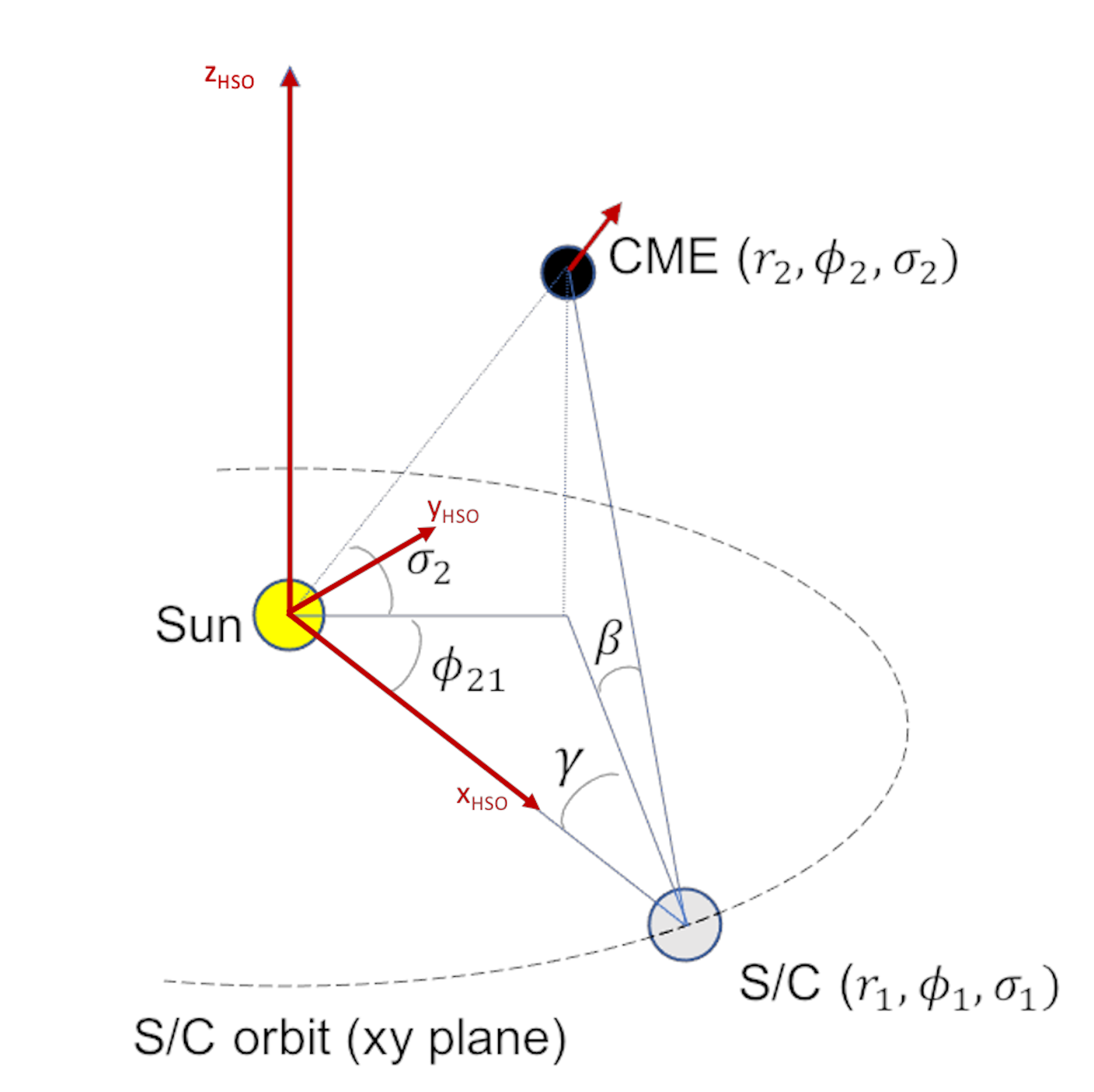}
\caption{CME front position $(r_{2}[t],\phi_{2},\sigma_{2})$ determination using spacecraft observations (PSP or STEREO-A) from coordinates $(r_1,\phi_{1},0)$. We determine the CME kinematics extracting the angles $\beta$ and $\gamma$, knowing the spacecraft position, and assuming some constraints for the CME (such as fixed angle or constant speed). Because of space constraints, we display $\phi_{2}-\phi_{1}$ as $\phi_{21}$. 
\label{fig:geometry}}
\end{figure}

We use the methodology developed for PSP observations by \cite{Liewer2019} as a starting point. It considers the CME as a point-like feature and the geometry shown in Figure \ref{fig:geometry} is used to determine the three-dimensional position as a function of time. From each available observation, we extract the angles $\beta_{obs}$ and $\gamma_{obs}$ in a camera FOV. $\beta_{obs}$ is the angle out of the orbit plane and $\gamma_{obs}$ resembles the elongation angle typically used on J-maps built from HI-1 and HI-2 observations. Both angles have vertex on the observing spacecraft. 

For each time $t$ in a set of consecutive images, we select a pixel with the CME feature of interest, normally the outermost point in a front-like structure. This pixel corresponds to the center of each small black square in Figures \ref{fig:cme1} and \ref{fig:cme2}. 
We don’t select the outermost part of the CME \#2 because its identification is not clear in the later images. We choose a parcel easy to track in the entire image set. This approach follows \citet{Liewer2019}, which tracked a relatively central point rather than the outermost CME feature.
From the pixel coordinates, we calculate $\beta_{obs}$ and $\gamma_{obs}$ using the routine \texttt{wispr\textunderscore camera\textunderscore coords.pro} \normalfont available on \texttt{SolarSoft} \normalfont \citep{Freeland1998} under the WISPR branch. This calculation accounts for the spacecraft location and attitude, as well as the camera projection and distortion effects. Since the spacecraft is moving, the transformation is time-dependent. Due to optical and projection effects and the changing scene, the spatial scale associated with each pixel depends on the FOV position. 

We also need to have the position of the spacecraft $(r_1[t],\phi_{1}[t],\sigma_{1}[t])$ at each time-instance $t$ in the orbit plane coordinates (Figure \ref{fig:geometry}). We define this heliocentric coordinate system with $x$ pointing towards the spacecraft and $y$ being located in the orbit plane towards the spacecraft ram direction. $\phi_1$ is the angle formed in the orbit plane with the $x$-axis and $\sigma_1$ is the angle between the orbit plane and spacecraft position. As the coordinate system here is pointing to the spacecraft at any $t$, we have $\phi_{1}[t]=0$ and $\sigma_{1}[t]=0$.

We consider that the tracked feature is a point-like structure at spherical coordinates $(r_{2}[t],\phi_{2}[t],\sigma_{2})$ where $r_2[t]$ is the radial distance to the Sun Center as a function of time $t$,  $\phi_2$ is the longitude and $\sigma_{2}$ is the latitude given in the spacecraft orbit coordinates system (see Figure \ref{fig:geometry} and Appendix \ref{sec:coordinate_conversion}). This structure has radial velocity $v_0$ at initial time-instance $t_0$ and acceleration $a$. Thus, we can write: 
\begin{equation}
r_2[t]=r_2[t_0] + v_0 [t-t_0]+a [t-t_0]^2/2
\label{eq:position}
\end{equation}
where $r_2[t_0]$ is the CME position at $t_0$.
In contrast to \citet{Liewer2019}, we do not consider the longitudinal angle $\phi_{2}$ to be fixed in time: 
\begin{equation}
\phi_{2}[t]=\phi_{2}[t_{0}] + \dot{\phi_{2}} [t-t_{0}]
\label{eq:deflection}
\end{equation}
where $\dot{\phi_{2}}$ is $\phi_2$ rate of change over time, i.e., the CME longitudinal deflection. Positive values show an increasing angle between the Sun-CME and Sun-spacecraft directions. This corresponds to a westward deflection for both PSP and STEREO-A.  

Following \cite{Liewer2019}, we relate the coordinates of the CME front with the projected angular coordinates ($\beta$, $\gamma$) in the image plane:

\begin{equation}
tan\,(\beta[t]) = \frac{tan\,(\sigma_{2})\ sin\,(\gamma[t])}{sin\,(\phi_2[t]-\phi_1[t])}
\label{eq:tanbeta}
\end{equation}

\begin{equation}
cotan\,(\gamma[t]) = \frac{r_{1}[t]-r_{2}[t]\ cos\,(\sigma_{2})\ cos\,(\phi_{2}[t]-\phi_{1}[t])}{r_{2}[t]\ cos\,(\sigma_{2})\ sin\,(\phi_{2}[t]-\phi_{1}[t])}
\label{eq:cotangamma}
\end{equation}

The parameters to be derived by the fit are $r_{2}[t_{0}]$, $v_0$, $a$, $\phi_{2}[t_0]$, $\dot{\phi_{2}}$, and $\sigma_{2}$. Spacecraft coordinates $(r_1[t], \phi_{1}[t],\sigma_{1}[t])=(r_1[t], 0, 0)$ are known. We use an extensive range for all 6 parameters ($r_{2}[t_{0}]$, $v_0$, $a$, $\phi_{2}[t_0]$, $\dot{\phi_{2}}$, and $\sigma_{2}$) in Equations \ref{eq:tanbeta} and \ref{eq:cotangamma} and derive $\beta_{der}[t]$ and $\gamma_{der}[t]$. The subscript $der$ is used here to distinguish these two angles from those extracted from observations ($\beta_{obs}[t]$ and $\gamma_{obs}[t]$). Then, we choose the set of parameters that result in minimum residual $\sigma$:

\begin{equation}
\sigma = \sum_{t=1}^{n}(|\beta_{obs}[t]-\beta_{der}[t]|)/n+(|\gamma_{obs}[t]-\gamma_{der}[t]|)/n
\label{eq:position_with_acceleration}
\end{equation}

where $n$ is the number of measurements. 

In the end, we obtain CME position as a function of time in the spacecraft orbit coordinates ($(r_{2}[t],\phi_{2}[t],\sigma_{2})$).

To reduce the number of free parameters, we have also considered some special cases: (i) constant speed $a=0$ and fixed longitude $\dot{\phi_{2}}=0$. This has 4 free parameters  and corresponds to the geometric analysis done in \cite{Liewer2019}; (ii) constant speed $a=0$ and (iii) constant longitude $\dot{\phi_{2}}=0$. 

\subsection{The residual of the fit and error estimation for each parameter}
\label{sec:goodness_fit}

We extract 6 time profiles of $\beta_{obs}$ and $\gamma_{obs}$ by tracking the CME front 6 times. For each tracking attempt and fit case, we derive a set of CME kinematics parameters ($\phi_{2}[t_0]$, $\sigma_{2}$, $v_0$, $a$, $r_{2}(t_{0})$ and $\dot{\phi_{2}}$). For each fit case, we adopt the median of each parameter as our result and the standard deviation as an estimate of error. We follow the same procedure for all results reported here.

\begin{figure*}[t]
\includegraphics[width=\hsize]{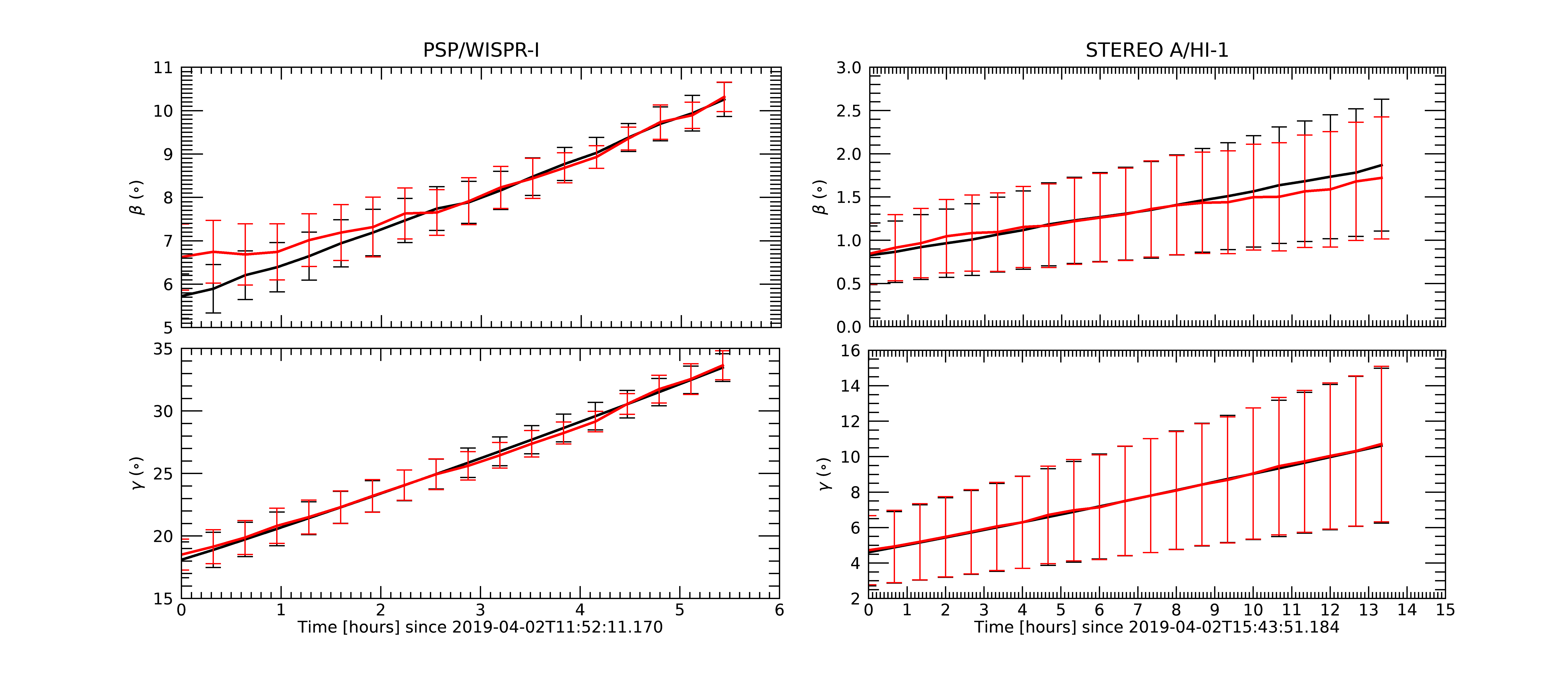}
\caption{Comparison of the observed and calculated $\beta$ and $\gamma$ parameters as a function of time for the CME \#2 using observations from WISPR-I (left panels) and HI-1 (right panels). The lines represent the median values, and the error bars correspond to the standard deviation calculated by repeating the fit with multiple sets of observations. The red lines indicate observations, and the black lines represent values calculated by the fit.}  
\label{fig:beta_gamma_hi_wi}
\end{figure*}

We show $\beta_{obs}$, $\gamma_{obs}$, $\beta_{der}$ and $\gamma_{der}$ for CME \#2 in Figure \ref{fig:beta_gamma_hi_wi} for a fit case using constant speed and no deflection. Overall, median values of $\beta_{der}$ and $\gamma_{der}$ (black lines) follow the observation ($\beta_{obs}$, $\gamma_{obs}$, red lines) both in WISPR-I and HI-1 observations. The highest discrepancy $|\beta_{obs}-\beta_{der}|$ is $\sim 1^{\circ}$ in the first WISPR-I measurement. The error bars for observed and derived angles have similar sizes, suggesting that the model reproduces the observations reasonably well.

One difference between the two instruments is in the angular range: $\beta_{obs}[t_n]-\beta_{obs}[t_1]$ and $\gamma_{obs}[t_n]-\gamma_{obs}[t_1]$ are both higher for WISPR-I (left panels in Figure \ref{fig:beta_gamma_hi_wi}). This can be explained by the wider WISPR-I FOV ($40^\circ$) when compared to HI-1 ($20^\circ$). Another reason can be the difference in spacecraft-CME distance, which is significantly smaller for PSP as the spacecraft is closer to the Sun and, therefore, to the CME.

We can also notice that the error bars in HI-1 observations (red bars in Figure \ref{fig:beta_gamma_hi_wi}, right panels) clearly increase with time. At the last time-instance, they are approximately 2 times higher than in the first. As the error bars result from visual inspections performed multiple times, and the CME is typically observed more clearly in the earlier measurements, $\beta_{obs}$ and $\gamma_{obs}$ change less from one visual inspection to another. On the other hand, this trend is not clear on WISPR-I observations used here. One possible explanation is that this is not a function of the spacecraft or instrument (as the results suggest) but rather of the particularities of the CME front, which can differ from one event to another.

For both CMEs and regardless of the set of free parameters used, we have fit residuals $\sigma < 0.1^{\circ}$ using observations from HI-1 and $\sigma<0.5^{\circ}$ from WISPR-I. This difference can probably be explained by the different ranges of $\beta$ and $\gamma$, which are proportionally higher in WISPR-I.

\subsection{Multiple spacecraft fit}
\label{sec:position_constrain}

In the fit done in Section \ref{sec:pos3d}, we calculate the CME kinematics independently for each imager. Thus, the CME front position $r_2$ can be different when comparing HI-1 and WISPR-I at a specific time. As there is no overlap with other CMEs, we explore the value of a fit that constrains the CME front position $r_2$ derived from one heliospheric imager to be the same for both.  

We use all measurements available: either when the CME is visible in one FOV only or both. $\dot{\phi_2}$ is no longer a free parameter. Hence we have only five: $r_{2}[t_{0}]$, $v$, $a$, $\phi_2$, $\sigma_2$. The first three parameters are identical in spacecraft-specific coordinates (Figure \ref{fig:geometry}), regardless of the spacecraft considered. $r_2[t]$ is constrained for both spacecraft. As in Section \ref{sec:pos3d}, here $r_2[t]$ is described by a second order fit (Equation \ref{eq:position}). The longitude and latitude parameters ($\phi_{2}$ and $\sigma_2$) are given in the spacecraft coordinate system, and we convert them from one spacecraft to the other, as described in Appendix \ref{sec:coordinate_conversion}.  

We use an extensive range for all 5 parameters ($r_{2}[t_{0}]$, $v_0$, $a$, $\phi_{2}[t_0]$ and $\sigma_{2}$) in Equations \ref{eq:tanbeta} and \ref{eq:cotangamma} and derive $\beta_{der}[t]$ and $\gamma_{der}[t]$. We calculate the residual of the fit $\sigma$ using the following equation:

\begin{equation}
\begin{split}
\sigma = \sum_{t=1}^{m}(|\beta_{obs}^{WISPR}[t]-\beta_{der}^{WISPR}[t]|/m+|\gamma_{obs}^{WISPR}[t]-\gamma_{der}^{WISPR}[t]|/m)\ +\\ 
\sum_{t=1}^{n}(|\beta_{obs}^{HI}[t]-\beta_{der}^{HI}[t]|/n+|\gamma_{obs}^{HI}[t]-\gamma_{der}^{HI}[t]|/n)
\end{split}
\label{eq:sigma_wi_hi}
\end{equation}
where $\beta_{der}^{WISPR}$, $\gamma_{der}^{WISPR}$, $\beta_{der}^{HI}$ and $\gamma_{der}^{HI}$ are the angular values in each spacecraft FOV calculated using Equations \ref{eq:tanbeta} and \ref{eq:cotangamma}, $n$ and $m$ are the number of measurements in WISPR-I and HI-1, respectively. Notice that this equation is the sum of the residuals associated with each spacecraft. 

One advantage of this methodology is that it increases the number of measurements available for the fit. This allows us to derive the CME kinematics over a longer period, when compared to single telescope measurements. We expect reduced fit errors as more measurements are available and the number of free parameters remains the same.

On the other hand, this method can lead to inconsistencies if the specific feature tracked by $\beta_{obs}$ and $\gamma_{obs}$ is not the same in both instruments. This can occur, for example, when we observe many CMEs in close timing. In this case, the association of the CME projection from one heliospheric imager to the other is ambiguous.

Hereafter, we refer to the fit introduced in this section as multiple spacecraft fit (MSF) and the cases which have observations of one spacecraft as single spacecraft fits (SSFs).

\section{Results and discussion}
\label{sec:results_and_discussion}

\subsection{CME \#1}
\label{sec:results_cme1}

CME \#1 was observed and tracked for over 6 hours in WISPR-I FOV and over 10 hours in HI-1 FOV. SSFs and MSF results are summarized in Tables \ref{tab:results_pos3d_cme1_multiple_methods} and \ref{tab:results_pos3d_cme1_acceleration_tied_position}, respectively. We illustrate the propagation direction and positions calculated using MSF in Figure \ref{fig:spacecraft_position_and_view}-a-b.  

In all cases, the results revealed a slow CME with speeds in the $200-300\ km\ s^{-1}$ range, propagating $5-8^{\circ}$ above the ecliptic plane, and between $47^{\circ}$ and $79^{\circ}$ eastward of the Sun-Earth line (HCI longitudes ranging from $36^{\circ}$ to $69^{\circ}$). 

For the SSF cases with acceleration $a$ as a free parameter (Table \ref{tab:results_pos3d_cme1_multiple_methods}, third column), we derived positive accelerations in both telescopes. However, this is a marginal result since the accelerations are of the same order as their errors. The MSF (Table \ref{tab:results_pos3d_cme1_acceleration_tied_position}) resulted in slightly stronger indication of acceleration ($0.556\pm0.351\ m\ s^{-2}$).  

\begin{figure}
\includegraphics[width=\hsize]{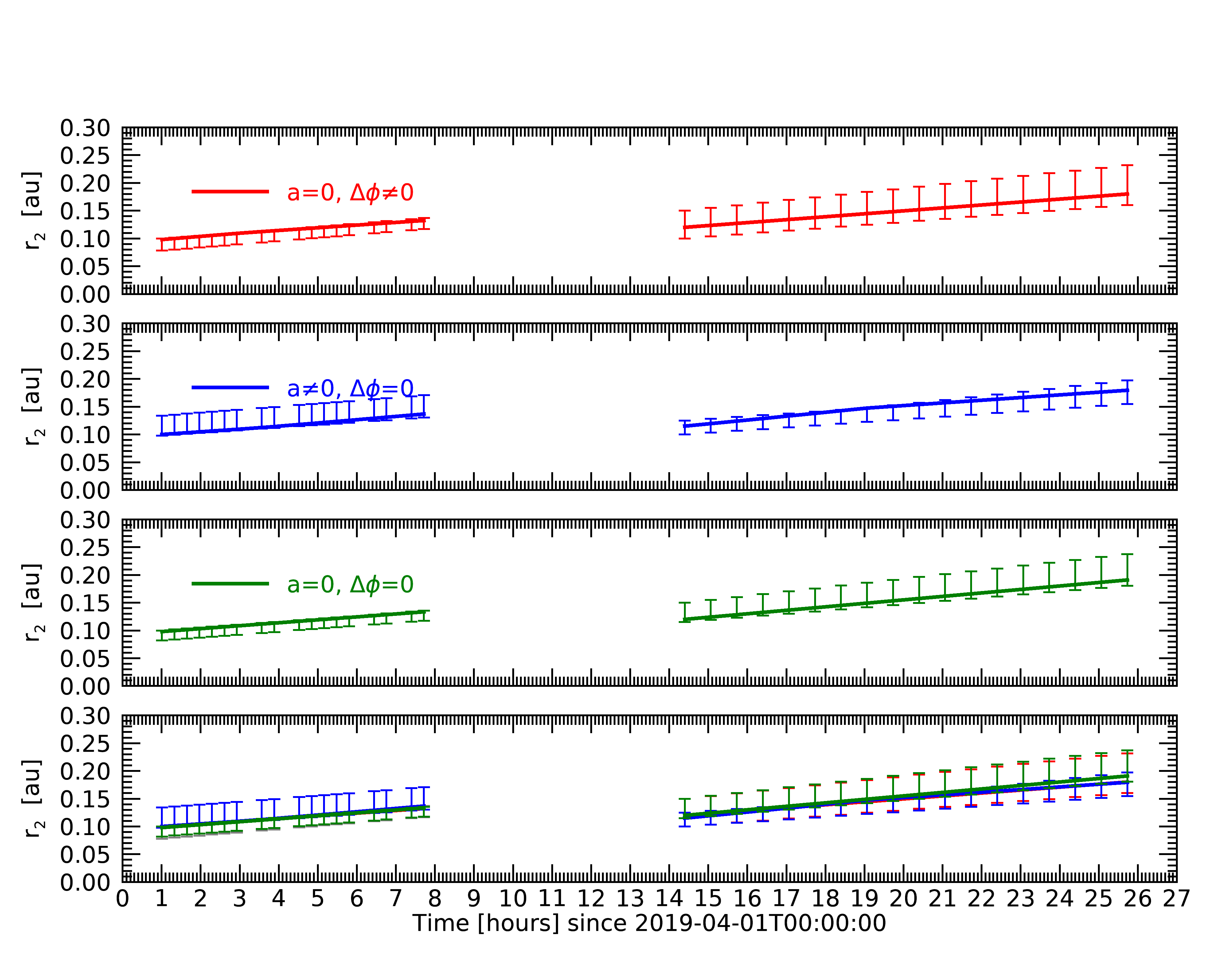}
\caption{CME \#1 position $r_2[t]$ derived by SSFs. The continuous lines show the median values, and the error bars indicate the minimum and maximum values. Each color represents a fit with a different set of free parameters, stated in each panel. We repeat all plots shown in the 3 upper panels is the bottom one to help to compare the fits. Results from MSF are not shown.}
\label{fig:cme1_r2_comparing_multiple}
\end{figure}

The CME front positions $r_2[t]$ derived using SSFs are shown in Figure \ref{fig:cme1_r2_comparing_multiple}. The 3 upper panels show the results under three different sets of free parameters. The bottom panel compares all three fits. For each instrument, $r_2[t]$ differences between the 3 cases are smaller than the error bars. So, the determination of the CME front distance is not significantly affected by the set of free parameters used, particularly acceleration and deflection that are not free parameters in all fits. We also found that the MSF $r_2$ agrees with those results from SSF, particularly in the first WISPR-I and last HI-1 measurements. 

To better evaluate the differences between WISPR-I and HI-1 SSFs, we extrapolated the WISPR-I $r_2$ 20 hours ahead to compare it against the HI-1 $r_2[t]$.  
In all cases, the extrapolated WISPR $r_2$ is $\sim 0.05\ au$ further away than the HI-1 $r_2$. This suggests that the $r_2$ discrepancy is because of the specific feature  tracked in HI-1 and WISPR-I rather than the assumptions for the fit, such as constant speed or propagation direction. In other words, the WISPR-I feature may be a distinct feature than the one in HI-1. There could be several reasons for this discrepancy. One reason may be that HI-1 is located further from and observes the event later than WISPR-I. Other possible reasons are the CME evolution and Thompson scattering effects.

In the case that considers $\dot{\phi_2}$ as a free parameter, the resulting deflections are marginal ($\dot{\phi_2}=-0.4\pm0.3^{\circ}\ h^{-1}$ for WISPR-I and $\dot{\phi_2}=0.4\pm0.5^{\circ}\ h^{-1}$ for HI-1). 

These $\dot{\phi_2}$ errors are unsurprising as the total deflection  $\Delta \dot{\phi_{2}} =\dot{\phi_{2}} (t_f-t_{0})$ (see Equation \ref{eq:deflection}) is smaller than the longitude error. For WISPR-I, $\Delta \dot{\phi_{2}}\sim-3^{\circ}$ while the longitude error is $10^{\circ}$. For HI-1, $\Delta \dot{\phi_{2}}\sim 5^{\circ}$ while the longitude error is $24^{\circ}$. More robust deflection estimates require fit over for longer periods. 

\begin{table*}
\begin{center}
\caption{Kinematics parameters derived for CME \#1 considering three different cases of free parameters (shown in columns).}
 \begin{tabular}{c c c c c} 
 \hline
  \hline
Parameter  & instrument & ($\dot{\phi_2}\neq 0$, $a=0$)  &  ($\dot{\phi_2}= 0,a\neq0$) & ($\dot{\phi_2}=0,a=0$)\\ 
\hline
\hline
\multirow{2}{*}{Maximum residual [$^{\circ}$]} & WISPR-I     & $0.3235$  & $0.2520$ & $0.3375$ \\
                                               & HI-1       & $0.0737$  & $0.0677$ & $0.0917$ \\
\hline
\multirow{2}{*}{Speed $t_0$ [$km\ s^{-1}$]} & WISPR-I               & $241\pm16$  & $210\pm13$  & $221\pm0$\\
                                            &  HI-1                 &  $241\pm31$ & $200\pm37$  & $261\pm35$\\
\hline
\multirow{2}{*}{Acceleration $[m\ s^{-2}]$} & WISPR-I          & -         & $0.556\pm1.303$ & - \\
                                            & HI-1             & -         & $1.667\pm1.488$ & - \\
\hline
\multirow{2}{*}{$\dot{\phi_{2}}$ [$^{\circ} h^{-1}$, HCI]} & WISPR-I    & $-0.4\pm0.3$  & -  & -\\
                                                      & HI-1       & $0.4\pm0.5$   & - & -\\
\hline
\multirow{2}{*}{Latitude at $t_0$  [$^{\circ}$, HCI]} &  WISPR-I   & $2\pm1$       & $1\pm1$  & $1\pm1$\\
                                                      & HI-1      & $-1\pm1$      & $-1\pm1$ & $-1\pm1$\\
\hline
\multirow{2}{*}{Longitude at $t_0$ [$^{\circ}$, HCI]} & WISPR-I    & $51\pm10$     & $36\pm7$ & $38\pm7$\\
                                                      & HI-1       & $68\pm24$     & $69\pm9$ & $68\pm9$\\
\hline
\multirow{2}{*}{Ecliptic latitude at $t_0$ [$^{\circ}$]}  & WISPR-I    & $8\pm2$       & $5\pm1$       & $5\pm1$\\
                                                            & HI-1       & $6\pm2$       & $6\pm1$       & $6\pm1$\\
\hline
\multirow{2}{*}{Angular distance from Earth at $t_0$ [$^{\circ}$]}  & WISPR-I    & $-64\pm10$    & $-79\pm7$    & $-77\pm7$\\
                                                                    & HI-1       & $-47\pm24$    & $-47\pm9$     & $-48\pm9$\\
 \hline
 \hline
\end{tabular}
\label{tab:results_pos3d_cme1_multiple_methods}
\end{center}
\end{table*}

\begin{table}
\begin{center}
\caption{Kinematics parameters derived for CME \#1 using fit with observations from both spacecraft. Acceleration is a free parameter here, but deflection ($\dot{\phi_2}= 0 $) isn't.} 
 \begin{tabular}{c c} 
  \hline
  \hline
 Parameter  & Value\\  
 \hline
 \hline
First observation ($t_0$)  & 2019-04-01T01:00:33 \\
Last observation ($t_f$)  & 2019-04-02T01:43:51 \\
Maximum residual [$^{\circ}$] & $   0.2907$ \\
Speed $t_0$ [$km\ s^{-1}$] & $         225\pm          10$ \\
Speed $t_f$ [$km\ s^{-1}$] & $         274\pm          24$ \\
Acceleration [$m\ s^{-2}$] & $    0.556\pm    0.351$ \\
Position at $t_0$ [au] & $    0.080\pm    0.000$ \\
Position at $t_f$ [au] & $    0.229\pm    0.006$ \\
Latitude [$^{\circ}$, HCI]  & $           2\pm           0$ \\
Longitude [$^{\circ}$, HCI]  & $          52\pm           1$ \\
Ecliptic latitude [$^{\circ}$]  & $           8\pm           0$ \\
Angular distance from Earth [$^{\circ}$]  & $         -63\pm           1$ \\
 \hline
 \hline
\end{tabular}
\label{tab:results_pos3d_cme1_acceleration_tied_position}
\end{center}
\end{table}

\subsubsection{Source region}

\begin{figure}
\includegraphics[width=\hsize]{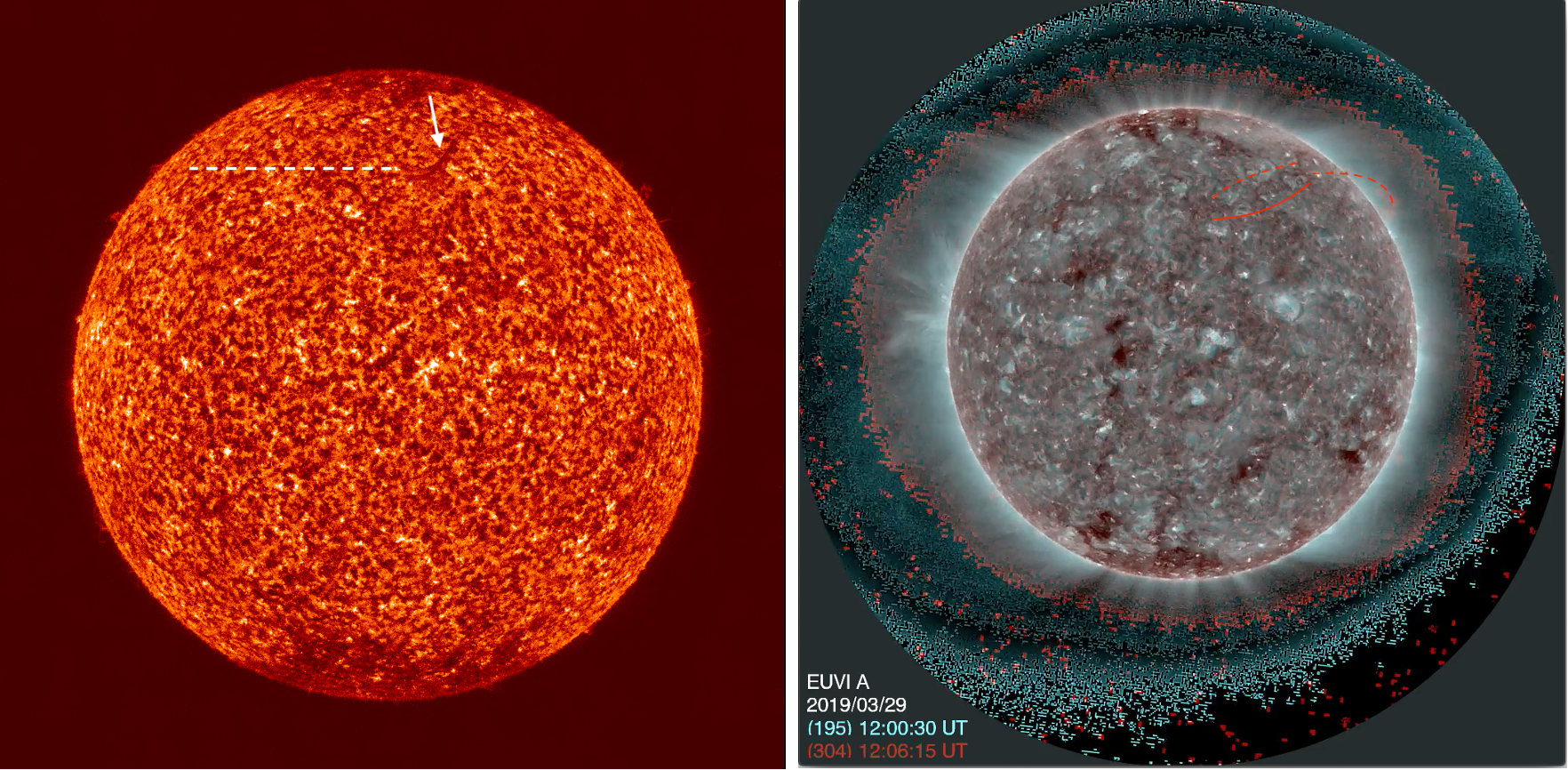}
\caption{CME \#1 source region in AIA 304\AA\ (left) and EUVI-A 195\AA /304\AA\  composite (right).}
\label{fig:source}
\end{figure}

To check whether these results were plausible, we searched for the source region of CME \#1. Given the low speed and solar minimum phase, it was unsurprising to see that there were no obvious low corona signatures associated with this CME. This event was, in other words, a 'stealth' CME \citep{robbrecht2009}. In our experience, however, a careful inspection of EUV images at different wavelengths and viewpoints can usually reveal the source regions, via the identification of tell-tale signs, such as widespread heating fronts or distant brightenings and slow rising loops systems. In this case, we identified the source region as an extended filament channel in the northwestern quadrant of the AIA (Atmospheric Imaging Assembly from Solar Dynamics Observatory - SDO) image (Figure~\ref{fig:source}). Because of the slow evolution of the eruption, it is difficult to pinpoint the exact time, but the rise started on March 29, 2019. On April 1, 2019, we detected diffuse brightentings along the channel in STEREO-A Extreme UltraViolet Imager (EUVI) images extending from $20-40^{\circ}$ longitude, corresponding to $38^{\circ}-58^{\circ}$ HCI longitudes. The close agreement between the CME \#1 and source longitudes gives us confidence in our measurement technique.

\subsection{CME \#2}
\label{sec:results_cme2}

CME \#2 is observed for approximately 16 hours when combining observations from WISPR-I and HI-1. Results from the 3 SSF cases (($\dot{\phi_2}\neq 0$, $a=0$), ($\dot{\phi_2}= 0,a\neq0$) and ($\dot{\phi_2}=0,a=0$)) are shown in Table \ref{tab:results_pos3d_cme2_multiple_methods}. For this CME, we added another variation of the fit using both instruments but only in the 4-hour period of overlapping observations. These results are shown in Table \ref{tab:results_pos3d_cme2_multiple_methods} in rows indicated by \say{both}. MSF results are shown in Table \ref{tab:results_pos3d_cme2_acceleration_tied_position} and illustrated in Figure \ref{fig:spacecraft_position_and_view} (panels \say{c} and \say{d}).

We found speeds of $\sim 300\ km\ s^{-1}$ with differences among the SSF cases mostly below $50\ km\ s^{-1}$. Regardless of the SSF, the HI-1 speed is higher than the  WISPR-I speed. The difference ranges from $20\ km\ s^{-1}$ to $60\ km\ s^{-1}$. Since the HI-1 observations are later than the WISPR-I observations, we interpret this as a possible positive acceleration. The MSF also suggests acceleration, with speeds increasing $157\ km\ s^{-1}$ between the first and last observations, which are approximately 17 hours apart.

Moreover, we derived significant accelerations  for WISPR-I SSF ($6.667\pm2.714\ m\ s^{-2}$) and MSF ($2.778\pm 1.152\ m\ s^{-2}$). We believe the differences in the acceleration magnitude can be explained by timing: the last time-instance $t_f$ used on MSF is $\sim 18$ hours later than in the WISPR-I SSF.  

For all fit cases, the propagation direction is above the ecliptic plane ($\sim 10^{\circ}$) and eastward from Earth (westward from STEREO-A). The HEE longitudes are $-21^{\circ}$ to $-61^{\circ}$, depending on the SSF fit. It is interesting to note that the latitude and longitude errors ($4^{\circ}$ and $1^{\circ}$, respectively) are clearly lower in the fit with simultaneous observations (rows labelled \say{both} in Table \ref{tab:results_pos3d_cme2_multiple_methods}) and in the MSF (Table \ref{tab:results_pos3d_cme2_acceleration_tied_position}). 

Regardless of the fit, the CME latitude and longitude derived from WISPR-I observations are higher than those derived from HI-1 (Table \ref{tab:results_pos3d_cme2_multiple_methods}). A likely reason for this discrepancy is the specific feature tracked. Although we tried to identify the same feature on both instruments, the association is subjective since the CME projection is each image is considerably different, particularly because of the different distances from the observing spacecraft, as we discussed for CME \#1. 

The longitudinal deflection fit indicates that CME \#2 is deflecting westward. We found $\dot{\phi_2}$ to be equal to $0.4-0.8^{\circ}/h$ with errors of $0.2-0.3^{\circ}/h$. This deflection magnitude is similar to the solar rotation speed ($\sim0.5^{\circ}/h$) and indicates that the CME is co-rotating with the Sun. 

\begin{table*}
\begin{center}
\caption{CME \#2 derived kinematics considering three different cases of free parameters (shown in columns).}
 \begin{tabular}{c c c c c } 
 \hline
 \hline
Parameter  & instrument & ($\dot{\phi_2}\neq 0$, $a=0$) &   ($\dot{\phi_2}= 0,a\neq0$) & ($\dot{\phi_2}=0,a=0$)\\ 
\hline
\hline
\multirow{3}{*}{Maximum residual [$^{\circ}$]} & WISPR-I      & $0.2414$  & $0.2874$ & $0.2875$ \\
                                               & HI-1         & $0.0581$  & $0.0727$ & $0.0862$ \\
                                               & Both         & $0.4334$  & $0.4277$ & - \\
                                                \hline
\multirow{3}{*}{Speed at $t_0$ [$km\ s^{-1}$]} & WISPR-I                & $281\pm33$ & $270\pm17$  & $321\pm20$\\
                                            & HI-1                  & $301\pm25$ & $320\pm46$  & $381\pm13$\\
                                            & Both                  & $280\pm47$ & $260\pm82$  & -\\
                                             \hline
\multirow{3}{*}{Acceleration $[m\ s^{-2}]$} &  WISPR-I           & -          & $6.667\pm2.714$         & - \\
                                            & HI-1              & -          & $0.556\pm0.978$         & - \\
                                            & Both              & -          & $13.333\pm52.317$       & - \\
                                            \hline
\multirow{3}{*}{$\dot{\phi_{2}}$ [$^{\circ}\ h^{-1}$, HCI]} &  WISPR-I    & $0.4\pm0.3$  & -  & -\\
& HI-1       & $0.8\pm0.2$   & - & -\\
& Both       & $0.8\pm0.8$   & - & -\\
\hline
\multirow{3}{*}{Latitude at $t_0$  [$^{\circ}$, HCI]} &   WISPR-I   & $8\pm3$       & $12\pm2$  & $12\pm1$\\
&  HI-1      & $0\pm1$       & $1\pm1$   & $0\pm0$\\
&  Both      & $9\pm1$       & $9\pm1$   & -\\
\hline
\multirow{3}{*}{Longitude at $t_0$ [$^{\circ}$, HCI]} &  WISPR-I    & $69\pm14$     & $95\pm10$ & $89\pm5$\\
& HI-1       & $56\pm8$      & $76\pm10$ & $59\pm1$\\
& Both       & $74\pm4$      & $74\pm4$  & -\\
\hline
\multirow{3}{*}{Ecliptic latitude at $t_0$ [$^{\circ}$]}  & WISPR-I    & $15\pm4$       & $19\pm2$         & $19\pm1$\\
&  HI-1       & $6\pm1$        & $8\pm1$          & $6\pm0$\\
&  Both       & $16\pm1$       & $16\pm1$         & -\\
\hline
\multirow{3}{*}{Angular distance from Earth at $t_0$ [$^{\circ}$]} &   WISPR-I    & $-48\pm14$     & $-21\pm10$       & $-27\pm5$\\
&  HI-1       & $-61\pm8$      & $-41\pm10$       & $-58\pm1$\\
&  Both       & $-43\pm4$      & $-43\pm4$        & -\\
\hline
\hline
\end{tabular}
\label{tab:results_pos3d_cme2_multiple_methods}
\end{center}
\end{table*}

\begin{table}
\begin{center}
\caption{CME \#2 kinematics derived with observations from both spacecraft. Acceleration is as a free parameter here, but deflection ($\dot{\phi_2}= 0 $) isn't.} 
 \begin{tabular}{c c} 
  \hline
  \hline
 Parameter  & Value\\  
 \hline
 \hline
First observation ($t_0$) & 2019-04-02T12:30:31 \\
Last observation ($t_f$)  & 2019-04-03T05:03:51 \\
Maximum residual [$^{\circ}$] & $   0.2085$ \\
Speed $t_0$ [$km\ s^{-1}$] & $         275\pm          30$ \\
Speed $t_f$ [$km\ s^{-1}$] & $         432\pm          51$ \\
Acceleration $[m\ s^{-2}]$ & $    2.778\pm    1.152$ \\
Position at $t_0$ [au] & $    0.070\pm    0.004$ \\
Position at $t_f$ [au] & $    0.213\pm    0.009$ \\
Latitude [$^{\circ}$, HCI]  & $           9\pm           1$ \\
Longitude [$^{\circ}$, HCI]  & $          76\pm           3$ \\
Ecliptic latitude [$^{\circ}$]  & $          16\pm           1$ \\
Angular distance from Earth [$^{\circ}$]  & $         -41\pm           3$ \\
 \hline
 \hline
\end{tabular}
\label{tab:results_pos3d_cme2_acceleration_tied_position}
\end{center}
\end{table}

\begin{figure}
\includegraphics[width=\hsize]{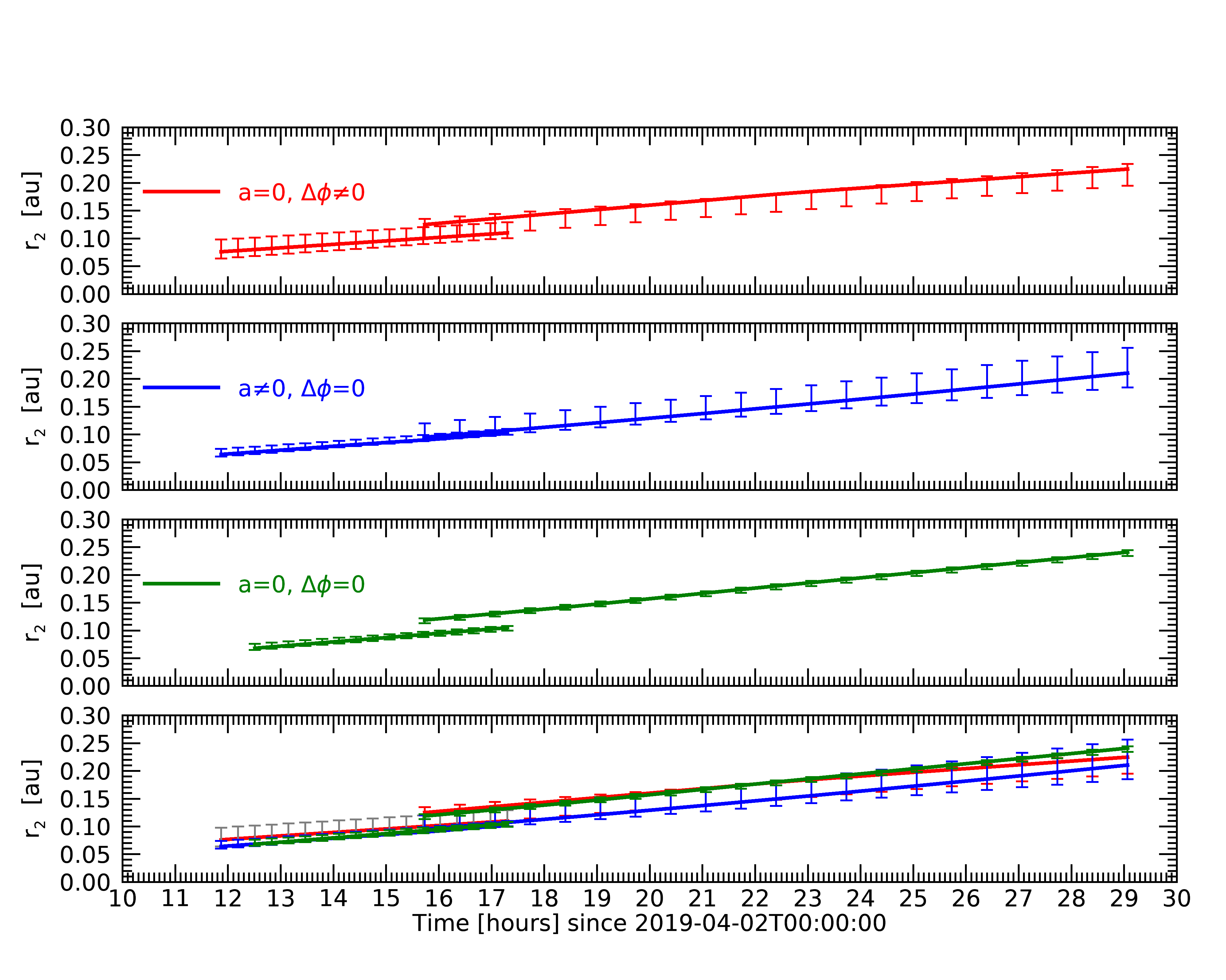}
\caption{CME \#2 position $r_2[t]$ derived by SSFs. The continuous lines show the median values, and the error bars indicate the minimum and maximum values. Each color represents a fit with a different set of free parameters, stated in each panel. We repeat all plots shown in the 3 upper panels in the bottom panel to help to compare the fits. Results from MSF are not shown.}
\label{fig:cme2_r2_comparing_multiple}
\end{figure}

We compare the CME position $r_2[t]$ in WISPR-I and HI-1 in Figure \ref{fig:cme2_r2_comparing_multiple}. In the period with simultaneous observations, $r_2[t]$ from WISPR-I is similar to $r_2[t]$ from HI-1 only in the SSF with acceleration as a free parameter (Figure \ref{fig:cme2_r2_comparing_multiple}, third panel from top). This suggests that the error of this fit is quite small as $r_2$ is calculated independently for each spacecraft. In the remaining cases, results are considerably different. HI-1 position is further away from the Sun $\sim 0.02\ au$ (red and green curves), which is similar or higher than the error bars.  When we compare all SSFs in each instrument FOV (bottom panel in Figure \ref{fig:cme2_r2_comparing_multiple}), the maximum difference in the position is $\sim 0.02\ au$ for WISPR-I and $\sim 0.05$ for HI-1. 

 \subsubsection{Previous studies and source region}
\label{sec:comparing }

The kinematics of CME \#2 using WISPR-I observations are also reported in \citet{Liewer2020}. The authors used a methodology based on \citet{Liewer2019}, which considers the same geometrical determination of the CME front shown in Figure \ref{fig:geometry} and resembles the SSF case without acceleration or deflection. One major difference with our study is that we tracked the CME outermost point in the front, while \citet{Liewer2020} tracked the lower dark “eye” of the skull-like flux rope, which is located closer to the Sun in a more central CME region. Thus, differences in the results are expected, as different features within the same CME may exhibit different motion patterns. 

\citet{Liewer2020} found the CME position to be $\sim 0.06\ au$ on April 2, 2019 at 12:09 UT. The position we derived for the same time is further away from the Sun (see Figure \ref{fig:cme1_r2_comparing_multiple}). This difference is consistent with the points tracked, since we are following a CME feature that is further away from the Sun. 

The HCI longitude ranges from $56^{\circ}$ to $95^{\circ}$, which encompasses the result from \citet{Liewer2020}  ($=67\pm1^{\circ}$). The same HCI longitude was found by \citet{Liewer2020} for the source region of this CME, identified to be AR12737, at its eruption time (March 31, 2019 at approximately 13 UTC).

The same holds for the speeds. Our results ($260\ km\ s^{-1}$ to $381\ km\ s^{-1}$) encompass the \citet{Liewer2020} result ($333\pm1\ km\ s^{-1}$). 

One major difference between our results and \citet{Liewer2020} is the error range that each parameter has. Regarding speed, the errors found here range from $17\ km\ s^{-1}$ to $33\ km\ s^{-1}$ considering only the results from WISPR-I observations, and from $13\ km\ s^{-1}$ to $82\ km\ s^{-1}$ considering both spacecraft. As for longitude, our errors range from $1^{\circ}$ to $14^{\circ}$, while in \citet{Liewer2020} the longitudinal error is $1^{\circ}$. We believe that the way we calculated this error explains this difference. We have 6 time-series of $(\beta_{obs}[t],\gamma_{obs}[t])$ derived from 6 independent trackings. In each time-series, we calculated the kinematics using our fit method. We then derive the error as 1-$\sigma$ between the 6 sets of kinematic parameters. \citet{Liewer2020} derived $\beta$ and $\gamma$ measurements from three independent trackings, and then calculated the mean value for each time. Using the mean time-series mean as input, the CME kinematic parameters (speed, position, longitude and latitude) were found using Levenberg-Marquardt least-squares. The uncertainty considered is 1-$\sigma$ error from the fit procedure. Overall, our results are consistent with those from \citet{Liewer2020}.

\subsection{How do errors compare between WISPR-I and HI-1?}

The SSF results from WISPR-I have lower errors for CME \#1 but higher errors for CME \#2. This trend is the same for all parameters studied here: distance, speed, acceleration, latitude, longitude and deflection. The only exception is the CME \#2 speed in one SSF case.

We calculated the errors from multiple visual identification of the CME front, i.e., from the different $\beta_{obs}[t]$ and $\gamma_{obs}[t]$ used in each fit. As the CME front identification is somehow subjective, we believe that the differences between WISPR-I and HI-1 results are due to the feature selection.

\subsection{Comparing fit with multiple and single spacecraft observations} 
\label{sec:comparing_fit_multiple_single}

In this section, we compare MSFs (Tables \ref{tab:results_pos3d_cme1_acceleration_tied_position} and \ref{tab:results_pos3d_cme2_acceleration_tied_position}) with SSFs (Tables \ref{tab:results_pos3d_cme1_multiple_methods} and \ref{tab:results_pos3d_cme2_multiple_methods}). For the latter, we choose the case that has acceleration as a free parameter but does not consider deflection ($a \ne 0, \dot{\phi_2}= 0 $), as this case has the same free parameters that the MSF.

For both CMEs, the discrepancies in distance $r_2$ and speed at $t_0$ are smaller than the errors. On the other hand, discrepancies exceed the errors for longitudes, CME \#2 acceleration and HI-1 latitudes. We also noticed that when we compared HI-1 and WISPR-I SSFs, the same parameters have discrepancies that exceed the errors bars. Thus, MSF particularities, such as the position constraint from one spacecraft to the other, are unlikely to be the major explanation for the discrepancies, as we also observed them when we compare SSFs from different instruments.

Regarding errors, the MSF errors have lower values for many parameters: longitude ($1-3^\circ$ when compared to $4-10^\circ$ from SSFs), $r_2[t_f]$ ($\sim0.01\ au$ rather than $\sim 0.05\ au$) and speed at $t_0$ ($10-40\ km\ s^{-1}$ when compared to $13-82\ km\ s^{-1}$). There is no clear trend for latitude or acceleration. 

The lower errors of MSF kinematic parameters possibly result from the higher number of measurements available: 38 for CME \#1 and 37 for CME \#2. Notice that MSF measurements correspond to the sum of those available on HI-1 and WISPR-I SSFs ($m+n$ on Equation \ref{eq:sigma_wi_hi}).

\subsection{Can the Thomson sphere explain the differences in longitude?}
\label{sec:thomson_sphere}

The tangent to the Sun of an observer's line of sight (LOS) for a particular elongation lies on the Thomson sphere (TS) \citep{Vourlidas2006}. The tangent point marks the location where the Thomson scattered emission from the free CME electrons in the corona is maximum along that LOS. This diameter of the sphere is equal to the Sun-observer distance. Features contribute progressively less to the LOS emission as their angular distance from the TS increases \citep[see details in][]{Vourlidas2006}.  

Because of the PSP's proximity to the Sun, the TS diameter for the WISPR observations is about 4 times smaller than for STEREO-A/HI-1. Since the two telescopes were approximately radially aligned when they observed the two CMEs, WISPR would be more sensitive to material closer to the spacecraft and hence would image the eastward (relative to Earth) flanks of the CME. Also, the WISPR LOS is much narrower than the HI-1 LOS and hence more sensitive to small-scale structures. Thus, we expect the CME propagation direction derived from WISPR observations to be eastward of the HI-1 estimates.
The longitudinal differences between WISPR and HI-1 for CME \#1 (all 3 cases of free parameters, Table \ref{tab:results_pos3d_cme1_multiple_methods}) are in agreement with this expectation. The opposite trend for CME \#2 (Table \ref{tab:results_pos3d_cme2_multiple_methods}). 

Errors in the longitudinal fits are in opposite directions in the two CMEs studied here. This suggests that the TS effect is not the dominant source of longitude discrepancies for CME \#2. With the derived kinematics, though, it is not possible to state which is the primary reason for CME \#1 discrepancy. The TS effect could be a source. It is possible that the specific features tracked in each instrument may lie behind the differences in the derived longitude for CME \#2 and perhaps for CME \#1.

\subsection{CME widths}
\label{sec:width}

The CME width is necessary for determining whether a CME intersects a location where in situ measurements are available. 
As a first order approximation, we considered the CME angular extent in latitude to be equal to the longitudinal one. As the latitude errors are lower ($\leq 3^{\circ}$) than the longitudinal ones ($\leq 24^{\circ}$), we estimated the latitudinal angular extent. To do so, we tracked the lowermost and utmost CME features as observed by WISPR-I, which correspond to the southern and northernmost CME portions. Here we repeated the same methodology used to calculate the CME front (Section \ref{sec:pos3d}) with the same set of WISPR-I images.

The angular extent we found here using WISPR-I observations are $13^{\circ}$ for CME \#1 and $16^{\circ}$ for CME \#2. To estimate the uncertainty of these angles, we repeated the fit using different sets of free parameters (Section \ref{sec:pos3d}) and 6 sets of CME features visual inspection in both the utmost and lowest CME features. We found a standard deviation of $3^{\circ}$.

We measured a $\sim20^{\circ}$ width for both CMEs in the HI-1 observations, which is a few degrees wider than WISPR-I measurements. We could not account for the difference so the CME is likely expanding or the longer HI1 LOS reveal fainter structure around the CME.

We realized that CME \#2 studied here is included in the HELCATS (Heliospheric Cataloguing, Analysis and Techniques Service) project (\url{www.helcats-fp7.eu/}). According to this catalog, the CME \#2 angular width is $20^{\circ}$, which is similar to our results.

\subsection{Exposure time}
\label{sec:exposure}

The exposure time is an important difference between the HI-1 and WISPR-I observations. It is $25.6 s$ for WISPR-I versus 20 minutes for HI-1, which translates to a radial motion of 0.5 solar radii (or 8 HI-1 pixels) for the derived speeds. This distance is much smaller than the position discrepancies (Figures \ref{fig:cme1_r2_comparing_multiple} and \ref{fig:cme2_r2_comparing_multiple}), which are approximately one order of magnitude larger ($\sim5$ solar radii).

The CME motion during the HI-1 exposure results in a positional uncertainty of $2^{\circ}$ in longitude and less than $1^{\circ}$ in latitude. These error estimates are based on the viewing angle, position, speed and propagation direction, and are smaller than the error estimates in Tables \ref{tab:results_pos3d_cme1_multiple_methods} and \ref{tab:results_pos3d_cme2_multiple_methods}.

Hence, the difference in exposure times is unlikely to significantly affect the derivation of positions in the two CMEs. However, the different exposure times could have important effects for fast CMEs. For example, CMEs with speeds above $1200\ km\ s^{-1}$ will travel more than 2 solar radii within an HI-1 exposure, which translates to over 30 HI-1 pixels.

\section{Are these CMEs observed in situ?}
\label{sec:geo}

We now check whether any of the CMEs was observed in situ, either in the Earth's vicinity (Section \ref{sec:insitu_wind}) or by PSP (Section \ref{sec:insitu_psp}). The objective is to see if the in situ measurements are consistent with the kinematics we derived in Sections \ref{sec:results_cme1} and \ref{sec:results_cme2}, particularly the propagation directions.

\subsection{Near Earth in situ observation}
\label{sec:insitu_wind}

\begin{figure*}
\centering
\includegraphics[width=12cm]{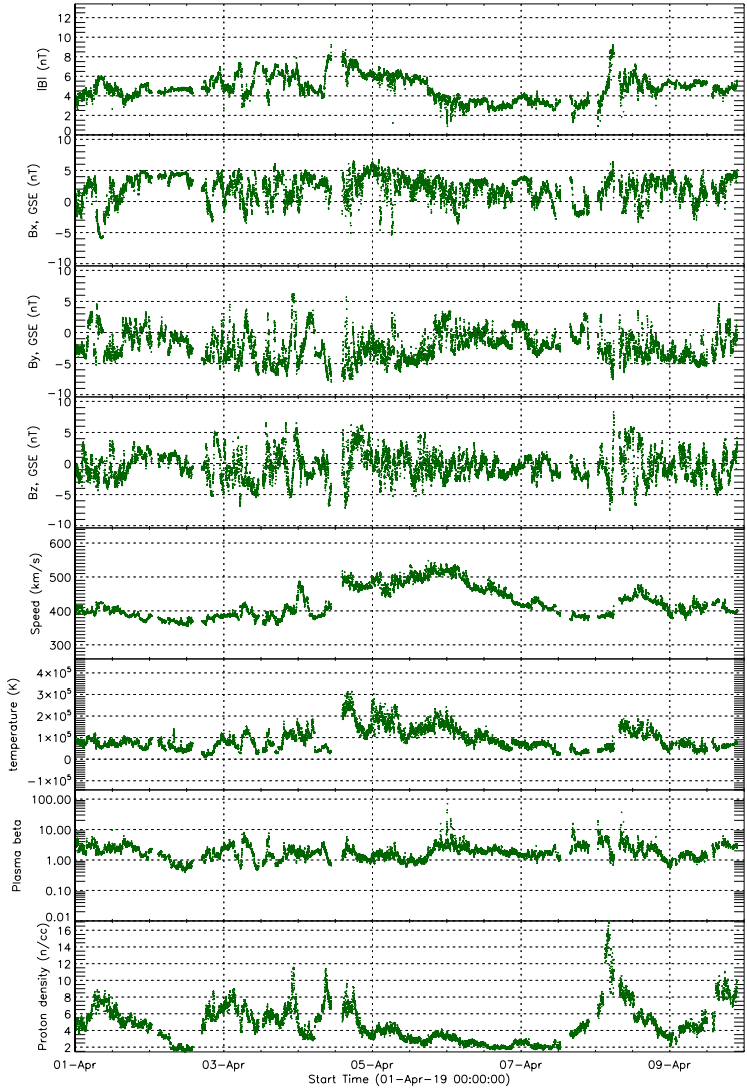}
\caption{Interplanetary magnetic field magnitude, orthogonal components and solar wind parameters (proton density, proton speed, temperature) and the plasma beta parameter observed near Earth.}
\label{fig:in_situ_wind_ace}
\end{figure*}

The interplanetary in situ magnetic field and solar wind parameters observed close to the Earth are shown in Figure \ref{fig:in_situ_wind_ace}. The panels, from top to bottom, show: (i) the magnetic field intensity; (ii-iv) its orthogonal components (given in geocentric earth-ecliptic coordinates - GSE); (v) the solar wind proton bulk speed; (vi) the proton temperature, (vii) the plasma beta parameter, which is the ratio between the plasma and magnetic pressures; and (viii) proton density. Data shown here is merged by \citet{King2005} from two spacecraft located close to the Earth: Advanced Composition Explorer \cite[ACE;][]{Stone1998}, particularly from magnetic field instrument \cite[MAG;][]{Smith1998} and Solar Wind Electron, Proton, and Alpha Monitor \cite[SWEPAM;][]{MCComas1998}, and Wind's Magnetic Field Instrument \cite[MFI;][]{Lepping1995} and Solar Wind Experiment \cite[SWE;][]{Ogilvie1995}.

The observed interplanetary signatures are inconsistent with the expected signatures from a typical CME: (i) the peak magnetic field intensities do not exceed 9 nT. This is among the lowest values observed \citep[see, e.g.,][]{Richardson2010}; (ii) the interplanetary magnetic field does not have clear rotation; (iii) the plasma beta parameter is higher than $1$ in almost any period after the peaks in the magnetic field intensity; 
(iv) the magnetic field variance does not decrease in the period with enhanced magnetic field. Normally higher fluctuations are observed in the period ahead of interplanetary CMEs. 

We can interpret these signatures as one of the following possibilities: (i) a disturbance possibly associated to the CME but not the flux rope itself, such as the sheath region; (ii) a flux rope portion far from its center, which does not produce a clear field rotation in the magnetic field components observed by a spacecraft crossing the structure; or (iii) a streaming interaction region, which is an interplanetary structure formed by the interaction of a slow wind stream followed by a fast one \citep[see, e.g.,][and references therein]{Richardson2018}. These regions are associated to increased interplanetary magnetic fields. 

This is in agreement with the CME propagation directions, which show both CMEs located eastward from the Sun-Earth line (Tables \ref{tab:results_pos3d_cme1_multiple_methods}, \ref{tab:results_pos3d_cme1_acceleration_tied_position}, \ref{tab:results_pos3d_cme2_multiple_methods} and \ref{tab:results_pos3d_cme2_acceleration_tied_position}). The CME longitudinal angle with the Sun-Earth line ranges from $47\pm24^{\circ}$ to $79\pm7^{\circ}$ for \#1 and from $21\pm10^{\circ}$ to $61\pm8^{\circ}$ for \#2, depending on the fit used. For both CMEs, the result for all fits indicate that the propagation direction of the CME longitude is eastward from the Earth (gray region in Figure \ref{fig:spacecraft_position_and_view}). Since the longitudinal width of the CMEs is rather small ($< 20^\circ$), it is unlikely they intercepted Earth.

\subsection{Are these CMEs observed in situ by PSP?}
\label{sec:insitu_psp}

CME \#1 front is in appropriate solar distance to reach PSP ($r_2 \sim 0.2\ au$) in the first hours of April 2, 2019 (Figures \ref{fig:spacecraft_position_and_view}-b and \ref{fig:cme1_r2_comparing_multiple}), when the spacecraft HCI longitude is $\sim 20^{\circ}$. As the CME \#1 HCI longitude is between $36\pm7^{\circ}$ and $69\pm9^{\circ}$, depending on the fit method considered, it is clearly westward from PSP and we do not expect it in situ. 

Doing similar analysis, we found that CME \#2 is also westward from PSP. This CME solar distance is equal to PSP's in the first hours of April 3, 2019 (Figures \ref{fig:spacecraft_position_and_view}-d and \ref{fig:cme2_r2_comparing_multiple}). At this time, the CME HCI longitude is between $56\pm8^{\circ}$ and $95\pm10^{\circ}$ and PSP's is $\sim 40^{\circ}$.

We also calculated the moment that PSP and each CME become longitudinally aligned. We used the longitude from MSF as the central longitude and the longitudinal width calculated in Section \ref{sec:width}. This takes place at April 3, 2019 12:00 UT for CME \#1 and at April 4, 2019 18:00 UT for CME \#2. We estimated an error of approximately 3 hours in the arrival time considering the width error (approximately $3^{\circ}$). Therefore, PSP is likely to have crossed the wakes of both CMEs. 

\begin{figure*}
\centering
\includegraphics[width=12cm]{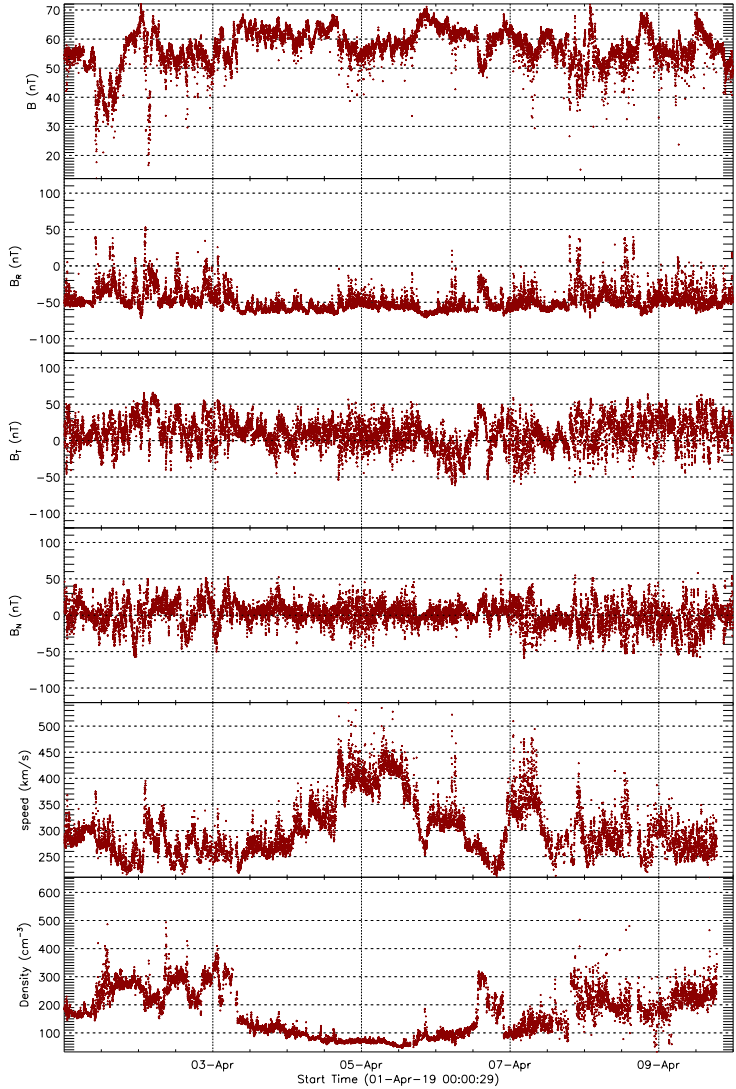}
\caption{Magnetic field magnitude, orthogonal components (radial, tangential and normal) and solar wind proton parameters (proton speed and density) observed in situ on PSP from April 1 to 10, 2019. The magnetic field and solar wind proton density are normalized by the square of PSP solar distance.}
\label{fig:in_situ_psp}
\end{figure*}

We show the PSP in situ measurements in Figure \ref{fig:in_situ_psp}. From top to bottom, the panels show the magnetic field intensity, its orthogonal components (radial, tangential and normal), the solar wind proton speed and density. Each panel show 1-minute medians. In the two bottom panels we use level 3 solar wind bulk parameters (density and speed) from the Solar Probe Cup (SPC), which is part of the Solar Wind Electrons Alphas and Protons (SWEAP) instrument suite \citep{Kasper2016, Case2020} onboard PSP. SPC is a Faraday Cup that looks directly to the Sun and measures ion and electron fluxes and flow angles as a function of energy. The measurements from this instrument are not obscured by PSP heat shield. The magnetic field data shown in the 4 upper panels is level 2 from FIELDS instrument \citep{Bale2016}. The magnetic field, its orthogonal components and the solar wind density are normalized by the square of the PSP distance. For all parameters, we use PSP position at April 1, 2019 0:00 UT as a reference point in the normalization.

The in situ measurements shown in Figure \ref{fig:in_situ_psp} are studied in \citet{Rouillard2020}, which interpreted the region with lower density observed from 2019 April 3 to 2019 April 7, except for a brief period in 2019 April 6, as PSP leaving the streamer belt flows. 

The interplanetary magnetic field lacks typical signatures of in situ measurements of CMEs, such as higher magnetic field with rotation and low plasma density \citep[see, e.g.,][]{Zurbuchen2006, Richardson2010}. Thus, the PSP in situ measurements agree with crossing behind the CMEs. Whether these observations reveal any information about the post-CME flow will require a more detailed study.

\section{Final remarks}
\label{sec:final_remarks}

We derive the kinematics of CMEs \#1 and \#2 (1 and 2 April, 2019) using  observations from two vantage points. The novelty here is that the observations were along similar angular distances from the CME but from very different heliocentric distances ($0.96\ au$ for HI-1, $0.19\ au$ for WISPR).  

We introduce a new method to derive the CME propagation using observations from both viewpoints as constraints.
This is based on the position determination introduced by \citet{Liewer2019}, which is suitable for the fast changes in the PSP longitude. This methodology adds CME acceleration and deflection as free parameters, along with speed, position, longitude and latitude. 
Considering HI-1 and WISPR-I observations (both combined and independently) and by doing a fit, we determined the CMEs kinematics in solar distances between $\sim 0.1\ au$ and $\sim 0.2\ au$.  

Our findings are summarized as follows:
\begin{enumerate}

\item Both CMEs are slow ($200-274\ km\ s^{-1}$ for \#1 and $260-432\ km\ s^{-1}$ for \#2) and propagate northward from the ecliptic plane ($5-8^\circ$ for \#1 and $6-19^\circ$ for \#2).

\item For CME \#2, the speed is higher on HI-1 FOV than on WISPR-I FOV. Depending on the SSF, the differences range from $20\ km\ s^{-1}$ to $60\ km\ s^{-1}$ and they are higher or equal than the speed errors estimated here. Since HI-1 observes the CME further from the Sun (start at $\sim 0.1\ au$),  CME \#2 may be accelerating. This agrees with SSFs with acceleration as a free parameter, which also resulted in positive values for this parameter. The MSF acceleration is also positive, with speed increasing from $275\ km\ s^{-1}$ to $432\ km\ s^{-1}$ while the CME propagates from $0.08\ au$ to $0.22\ au$. For CME \#1 we also found acceleration, but the magnitude is the same order as the error, so it is a marginal result.

\item Both CMEs propagate eastward from Earth and westward from PSP and STEREO-A. Depending on the fit case considered, their angular distance with Earth is between $47^{\circ}$ and $79^{\circ}$ for CME \#1, and between $21^{\circ}$ and $61^{\circ}$ for CME \#2. The longitudinal differences remain whether we compare SSFs cases with different free parameters or SSFs from different instruments. Thus, the discrepancies in the longitude are unlikely to be due to the instrument or set of free parameters. We understand that the difference in CME longitude from one method to the other is a combination of error from tracking different CME front features on HI-1 and WISPR-I. 

\item To assess the errors in our kinematic results, we visually tracked and fitted the CME front several times. We then used the spread in the resulting kinematic parameters to derive the error estimates of up to $\sim0.05\ au$ in radial distance, $\sim10^{\circ}$ in latitude and, $\sim24^{\circ}$ in longitude. These errors cannot fully explain the discrepancies in the position and direction of propagation found when comparing results from HI-1 and WISPR-I. The most likely explanation is that we are not tracking the same feature in both telescopes.

\item Fits with WISPR-I observations result in lower errors for CME \#1  but higher errors for CME \#2 when compared to those from HI-1. Thus, we do not see any clear trend of smaller fit errors in any instrument/spacecraft.

\item We allowed CME longitudinal deflection as a free parameter in the fit. For CME \#2, we found significant westward deflections for both WISPR-I and HI-1 SSFs. The longitudinal deflection magnitude is consistent with the solar rotation, and we believe that this is an indication that  CME \#2 is co-rotating with the Sun up to $\sim 0.2\ au$. The deflection estimates for CME \#1 are smaller than their errors.

\item  Thomson scattering considerations suggest that WISPR-I estimates should be based eastward of HI-1 estimates, if Thomson scattering effects were important. We found this trend for CME \#1 (1 April) but not for CME \#2 (2 April). For the latter, longitudes derived using WISPR-I observations are more than $13^{\circ}$ westward compared to HI-1 observations. A plausible explanation for this difference is that the specific features tracked is different in each instrument FOV and this affects the derived longitudes more than the Thomson scattering. 

\item The MSF has lower errors for position, speed and longitude. One possible explanation is the higher number of measurements available when compared to the SSF. For the remaining parameters, there is no clear trend.

\item Positions and speeds derived using the MSF agree with those derived using SSFs. Acceleration, longitude and latitude have discrepancies between corresponding HI-1 and WISPR-I SSFs and also when we compare MSF with a SSF.

\item We inspected in situ measurements from PSP and near Earth spacecraft and found no clear signatures of interplanetary counterparts of any CME studied here. This finding supports the CME propagation directions from both MSF and SSFs. 
\end{enumerate}

In future studies, we plan to apply the methodology discussed here to other CMEs observed both by PSP and STEREO-A, particularly to those with in situ measurements close to the Sun by PSP.

\begin{acknowledgements}
C.R.B. acknowledge the support from the NASA STEREO/SECCHI (NNG17PP27I) program. A.V. is supported by the WISPR Phase-E funding and NASA 80NSSC19K1261 grant. 

Parker Solar Probe was designed, built, and is now operated by the Johns Hopkins Applied Physics Laboratory as part of NASA’s Living with a Star (LWS) program (contract NNN06AA01C). Support from the LWS management and technical team has played a critical role in the success of the Parker Solar Probe mission.

The Wide-Field Imager for Parker Solar Probe (WISPR) instrument was designed, built, and is now operated by the US Naval Research Laboratory in collaboration with Johns Hopkins University/Applied Physics Laboratory, California Institute of Technology/Jet Propulsion Laboratory, University of Gottingen, Germany, Centre Spatiale de Liege, Belgium and University of Toulouse/Research Institute in Astrophysics and Planetology. WISPR data are available for download at \url{https://wispr.nrl.navy.mil/}.

We acknowledge the NASA Parker Solar Probe Mission and the SWEAP team led by J. Kasper for use of data. SWEAP data are available at \url{http://sweap.cfa.harvard.edu/}. The FIELDS experiment on the Parker Solar Probe spacecraft was designed and developed under NASA contract NNN06AA01C.  FIEDLS data can be downloaded at \url{https://fields.ssl.berkeley.edu/}

The Sun Earth Connection Coronal and Heliospheric Investigation (SECCHI) was produced by an international consortium of the Naval Research Laboratory (USA), Lockheed Martin Solar nd Astrophysics Lab (USA), NASA Goddard Space Flight Center (USA), Rutherford Appleton Laboratory (UK), University of Birmingham (UK), Max Planck Institute for Solar System Research (Germany), Centre Spatiale de Liége (Belgium), Institut d’Optique Theorique et Appliquée (France), and Institut d'Astrophysique Spatiale (France). STEREO/SECCHI data are available for download at \url{https://secchi.nrl.navy.mil/}.

We acknowledge use of NASA/GSFC's Space Physics Data Facility's CDAWeb service, and OMNI data. This data is available for download at \url{https://omniweb.sci.gsfc.nasa.gov/html/ow_data.htm}.

This research has made use of the Solar Wind Experiment (SWE) and Magnetic Field Investigations (MFI) instrument's data onboard WIND. We thank to the Wind team and the NASA/GSFC's Space Physics Data Facility's CDAWeb service to make the data available. Wind data are available from \url{https://cdaweb.sci.gsfc.nasa.gov}.

We thank P. C. Liewer and J. Qiu for useful discussions.

\end{acknowledgements}

%
   \bibliographystyle{aa} 
   \bibliography{sample63} 
%

\begin{appendix}

\section{Coordinate conversion from/to spacecraft orbit}
\label{sec:coordinate_conversion}

Examples of coordinate systems commonly adopted by Heliophysics community are the Heliocentric Earth Equatorial \citep[HEEQ;][]{Hapgood1992}, Heliocentric Earth Inertial \citep[HCI;][]{Burlaga1984}, and Carrington, which are based on the solar equatorial plane, and the Heliospheric Earth Ecliptic \citep[HEE;][]{Hapgood1992}, based on the ecliptic plane. For a complete explanation about all these coordinate systems, readers are referred to \citet{Thompson2006}. 

The coordinate system used here to derive the CME position (Figure \ref{fig:geometry}) is not frequently used. It is heliocentric, and its $x$ and $y$ axes are in the spacecraft orbit. Neither PSP nor STEREO orbits lie in nor are parallel to the solar equatorial nor ecliptic plane.

We call the coordinate system used in Figure \ref{fig:geometry} heliocentric spacecraft orbital (HSO). Its $x$ points from the Sun to the spacecraft position, $z$ is perpendicular to the orbit plane, and $y$ lies in the orbit plane and forms the orthogonal right-handed coordinate system. In the case of STEREO-A and PSP, whose orbits are both counter-clock wise from an observed located in the north, the $y$ corresponds to a future position in the orbit. Notice that this coordinate system is spacecraft specific and all axes change as the spacecraft orbits the Sun. For PSP, the $x$ and $y$ axes change tents of degrees in a matter of days, particularly close to the perihelium.

In order to compare the CME position  for PSP and STEREO-A we need to convert them to a common coordinate system, such as HEEQ or HEE. In both the $x$ axis follows the Earth's position around the Sun and they are not inertial for a given CME. Instead, we use the HCI coordinates, which do not follow the solar rotation nor the Earth's rotation around the Sun. The HCI $x$-axis is the solar ascending node on ecliptic of J2000.0, $z$ axis is perpendicular to the solar equatorial plane and $y$ is also in the solar equatorial plane and forms the right-handed coordinate system.

A given position vector $\vec{r}$ can be written in both HSO ($o$) and HCI ($i$) coordinate systems:
\begin{equation}
\vec{r}=r_{x_o} \hat{x_o} + r_{y_o} \hat{y_o} +r_{z_o} \hat{z_o}
\label{eq:r_in_terms_of_o}
\end{equation}
\begin{equation}
\vec{r}=r_{x_i} \hat{x_i} + r_{y_i} \hat{y_i} +r_{z_i} \hat{z_i}
\end{equation}
where $\hat{x_o},\hat{y_o},\hat{z_o}$ are unit vectors in the HSO coordinate system and $\hat{x_i},\hat{y_i},\hat{z_i}$ in the HCI.

Multiplying \ref{eq:r_in_terms_of_o} by $\hat{x_i}$, $\hat{y_i}$, $\hat{z_i}$ we get:

\begin{equation}
r_{x_i} = r_{x_o} \hat{x_o} \hat{x_i} + r_{y_o} \hat{y_o} \hat{x_i} + r_{z_o} \hat{z_o}\hat{x_i}
\end{equation}
\begin{equation}
r_{y_i} = r_{x_o} \hat{x_o} \hat{y_i} + r_{y_o} \hat{y_o} \hat{y_i} + r_{z_o} \hat{z_o}\hat{y_i}
\end{equation}
\begin{equation}
r_{z_i} = r_{x_o} \hat{x_o} \hat{z_i} + r_{y_o} \hat{y_o} \hat{z_i} + r_{z_o} \hat{z_o}\hat{z_i}
\end{equation}

which can be written in the following matrix form: 
\begin{equation}
\begin{bmatrix} r_{x_i} \\ r_{y_i} \\ r_{z_i} \end{bmatrix}=
\begin{bmatrix}  \hat{x_o} \hat{x_i} & \hat{y_o} \hat{x_i} & \hat{z_o} \hat{x_i} \\ \hat{x_o} \hat{y_i} & \hat{y_o} \hat{y_i} & \hat{z_o} \hat{y_i} \\ \hat{x_o}  \hat{z_i} & \hat{y_o} \hat{z_i} & \hat{z_o} \hat{z_i} \end{bmatrix}  \begin{bmatrix} r_{x_o} \\ r_{y_o} \\ r_{z_o} \end{bmatrix}
\label{eq:matrix_form}
\end{equation}

By definition, $\hat{x_o} \hat{x_i}=cos \phi_{x_i,x_o}$ where $\phi_{x_i,x_o}$ is the angle between the unit vectors. Notice that $\phi_{x_i,x_o}$ depends only on the two coordinate systems used and is identical for any vector used. The same conclusion can be derived for any elements of this matrix. This matrix transforms any vector from $o$ coordinates to $i$ coordinates and hereafter we refer it as $T_{o2i}$. Thus, we can rewrite \ref{eq:matrix_form} as:

\begin{equation} R_i=T_{o2i} R_o \end{equation}

Multiplying by the inverse matrix of $T_{o2i}$ both right and left-hand side:
\begin{equation} T_{o2i}^{-1} R_i=T_{o2i}^{-1} T_{o2i} R_o \end{equation}

Thus, we have
\begin{equation} R_o = T_{o2i}^{-1} R_i  \end{equation}

Hereafter we refer $T_{o2i}^{-1}$ as $T_{i2o}$.

We need to find the CME position in HCI coordinates $\vec{R_i}$ and we have $\vec{R_o}$. To do so, we need to know the transformation matrices $T_{o2i}$. 

Notice that since the HSO coordinate system is rotating to follow the spacecraft position and changes as a function of time, the transformation matrices $T_{o2i}$ and $T_{i2o}$ need to be calculated for each time instance and for each spacecraft. 

We can calculate the elements of the first row of $T_{o2i}$ ($\hat{x_o} \hat{x_i},  \hat{y_o} \hat{x_i} , \hat{z_o} \hat{x_i}$) by using $\vec{a}=\hat{x_o}$ where $\hat{x_o}$ is the unit vector pointing from the Sun to the spacecraft location, i.e., the $x$ axis in the HCO coordinate system. Thus, we can write:

\begin{equation}
\begin{bmatrix} a_{x_i} \\ a_{y_i} \\ a_{z_i} \end{bmatrix}=
\begin{bmatrix}  \hat{x_o} \hat{x_i} & \hat{y_o} \hat{x_i} & \hat{z_o} \hat{x_i} \\ \hat{x_o} \hat{y_i} & \hat{y_o} \hat{y_i} & \hat{z_o} \hat{y_i} \\ \hat{x_o}  \hat{z_i} & \hat{y_o} \hat{z_i} & \hat{z_o} \hat{z_i} \end{bmatrix}  \begin{bmatrix} 1 \\ 0 \\ 0 \end{bmatrix}.
\end{equation}

The HCI coordinates of $\vec{a}$ ($a_{x_i},a_{x_i},a_{x_i}$), which correspond to the unit vector pointing from the Sun to the spacecraft, are known and can be derived using \texttt{get\textunderscore sunspice\textunderscore coords.pro} \normalfont from SolarSoft.

Similarly, the terms in the second and third rows of $T_{o2i}$ can be determined by doing $\vec{a}=\hat{y_o}$ and $\vec{a}=\hat{z_o}$. HCI components of $\hat{z_o}$ can be derived doing by the cross product of $\hat{x_o}$ at two different measurements and $\hat{y_o}$ is found by doing $\hat{z_o} \wedge \hat{x_o}$.

By doing so, we have determined the transformation matrix $T_{o2i}$ and we can convert the CME position vector from HSO coordinates, which are derived in Section \ref{sec:pos3d} and are spacecraft and time specific, to HCI coordinates, which are inertial and can be easily converted to many other coordinate systems using  \texttt{convert\textunderscore sunspice\textunderscore coord.pro} \normalfont from SolarSoft. 

\end{appendix}

\end{document}